\documentclass{aa}  

\usepackage{graphicx}
\usepackage{txfonts}
\usepackage{lipsum}
\usepackage{subcaption}
\usepackage{lscape}
\usepackage{placeins}
                                
\usepackage[
 bookmarks=true,
 pdfnewwindow=true,
 colorlinks=true,
 linkcolor=blue,
 citecolor=blue,
 filecolor=blue,
 urlcolor=blue,
 final=true,
]{hyperref}
\usepackage[
 print-unity-mantissa=false,
 exponent-product=\ensuremath{\cdot},
 uncertainty-mode = separate,
]{siunitx}
\usepackage{amsmath}  % \eqref., Advanced maths commands
\usepackage{xcolor}

\DeclareSIUnit{\photon}{\text{\ensuremath{\textnormal{ph}}}}
\DeclareSIUnit{\erg}{\text{\ensuremath{\textnormal{erg}}}}
\DeclareSIUnit{\deg}{\text{\ensuremath{\textnormal{deg}}}}
\DeclareSIUnit{\gauss}{\text{\ensuremath{\textnormal{G}}}}
\DeclareSIUnit{\solarmass}{\text{\ensuremath{\textnormal{M}_\odot}}}
\DeclareSIUnit{\solarradius}{\text{\ensuremath{\textnormal{R}_\odot}}}
\DeclareSIUnit{\year}{\text{\ensuremath{\textnormal{yr}}}}
\DeclareSIUnit{\parsec}{\text{\ensuremath{\textnormal{pc}}}}
\DeclareSIUnit{\magnitude}{\text{\ensuremath{\textnormal{mag}}}}
\DeclareSIUnit\angstrom{\text {Å}}

\def \d {\mathrm{d}}

\begin{document}

   \title{Intra-day variability of the gamma-ray/optical emission of 3C~454.3: the complexity beyond the long-term correlation}
   %\subtitle{hidden complexity in a strong correlation}

    \titlerunning{Intra-day variability of the $\gamma$-ray/optical emission of 3C~454.3}

    \author{
        Gabriele Panebianco\inst{1}\corrauth{gabriele.panebianco@inaf.it}
        \and Marco Tavani\inst{2,3}\email{marco.tavani@inaf.it}
        \and Valerio Vittorini\inst{2}\email{valerio.vittorini@inaf.it}
        \and Paolo Coppi\inst{4}\email{paolo.coppi@yale.edu}
        \and Andrea Bulgarelli\inst{1}\email{andrea.bulgarelli@inaf.it}
        \and Claudia M. Raiteri\inst{5}\email{claudia.raiteri@inaf.it}
        \and Massimo Villata\inst{5}\email{massimo.villata@inaf.it}
    }
    \institute{
        INAF - Osservatorio di Astrofisica e Scienza dello spazio di Bologna (OAS), Via Piero Gobetti 93/3, 40129 Bologna, Italy
        \and INAF - Istituto di Astrofisica e Planetologia Spaziali (IAPS), Via del Fosso del Cavaliere 100, 00133 Roma, Italy
        \and Dipartimento di Fisica, Università di Roma Tor Vergata, via della Ricerca Scientifica 1, I-00133 Roma, Italy
        \and Department of Astronomy, Yale University, PO Box 208101, New Haven, CT 06520-8101, USA
        \and INAF - Osservatorio Astrofisico di Torino, Via Osservatorio 20, I-10025 Pino Torinese, Italy
    }
    
    \authorrunning{G. Panebianco et al.}

    \date{Received October 7, 2026; accepted November 18, 2026}

    %%%%%%%%%%%%%% ABSTRACT
\abstract
{Simultaneous $\gamma$-ray/optical observations provide a powerful probe of particle acceleration and radiative processes in blazar jets.
The blazar 3C~454.3 exhibits a significant long-term correlation between these bands, which has been interpreted as strong support for the canonical one-zone leptonic emission model.
However, 3C~454.3 also shows strong intra-day variability in both bands.
Whether this correlation also exists for individual flares has not been systematically studied.}
{We analyse the $\gamma$-ray/optical correlation and intra-day variability of 3C~454.3 over $17$ years on timescales $\leq \SI{24}{\hour}$, investigating how short-timescale behaviour relates to the long-term correlation.}
{For the time period covering the years 2007 to 2024, we analysed AGILE and Fermi-LAT observations, producing light curves with multiple binning from $3$ to $48$ h.
We matched them to simultaneous R-band photometry from WEBT and public optical data sets.
We looked in detail at the flux and spectral time evolution of individual flaring epochs, estimating the time-variable correlation properties and Compton dominance from the optical and $\gamma$-ray luminosities.}
{The long-term $\gamma$-ray/optical correlation is confirmed, but exhibits a large intrinsic dispersion and a correlation slope $\eta\simeq1.6-1.7$ that cannot be explained by a single canonical one-zone scenario.
We see correlation strengths ranging from weak to strong, correlation slopes spanning $\eta\simeq0.4-2.0$, orphan flares in both bands breaking the correlation, diverse flare morphologies, distinct Compton dominance evolution patterns with rapid excursions from values $\sim 10$ up to $120$, and heterogeneous spectral trends.
The shortest resolved variability ($t_{\rm var}\lesssim3$ h) also provides independent evidence that localized emission regions are required following causality arguments.}
{The well-known $\gamma$-ray/optical correlation of 3C~454.3 is not produced by a single emission regime, but emerges from the superposition of multiple epoch-dependent physical states.
Increasing the temporal resolution reveals progressively richer variability, highlighting the importance of systematic intra-day, multi-epoch studies for interpreting the emission of bright FSRQs.}


\maketitle
%%%%%%%%%%%%%%%%%%%%%%%%%%%%%%%%%%%%%%%%%%%%%%%%%%%%%%%%%%%%%%
% Line numbers every 1 line
%\modulolinenumbers[1]
\nolinenumbers

%%%%%%%%%%%%%%%%% BODY OF PAPER %%%%%%%%%%%%%%%%%%
\section{Introduction}
\label{sec:introduction}
% 1 - BLAZARS EMISSION MECHANISMS
Blazars are Active Galactic Nuclei (AGN) with relativistic plasma jets oriented close to the observer's line of sight \citep{urry_padovani_AGNUnification_1995, Padovani_AGNreview_2017A&ARv..25....2P}.
Their broadband spectral energy distribution (SED) features two distinct peaks.
In flat-spectrum radio quasars (FSRQs), these occur in the millimetre/near-infrared and MeV-GeV bands, respectively, whereas in BL Lacertae objects (BL Lacs) they span from infrared to X-rays and from X-rays up to TeV energies \citep{Ghisellini_BlazarSequence_2016Galax...4...36G,prandini_blazarreview_2022}.
The low-energy peak originates from synchrotron radiation of relativistic electrons.
The high-energy peak is attributed either to leptonic processes via inverse Compton (IC) scattering of synchrotron photons \citep[synchrotron self-Compton, SSC;][]{maraschi_SSC_1992ApJ...397L...5M, bloom_SSC_1996ApJ...461..657B} or external fields like the accretion disc, broad-line region (BLR), or dusty torus \citep[external Compton, EC;][]{dermer_EC_1993ApJ...416..458D, sikora_EC_1994ApJ...421..153S, ghisellini_EC_2009MNRAS.397..985G}, or to hadronic processes such as proton synchrotron radiation and secondary cascades \citep{muecke_hadronicflare_2001APh....15..121M, Dimitrakoudis_hadronicmodels_2012A&A...546A.120D, cerruti_jetprocesses_2020}.
Structurally, emission models are divided into one-zone scenarios, where emission is localized in a single compact region \citep{celotti_powerblazarjets_2008}, and multi-zone, which distribute emitting particles along the jet or in moving plasmoids \citep{villata_curved_multizone_1999A&A...347...30V, chen_multizone_SSC_2011MNRAS.416.2368C, marsher_multizone_turbulence_2014ApJ...780...87M, raiteri_cta102curved_2017Natur.552..374R, shukla_nature_3C279minutes_2020}.

% 2 - BLAZAR VARIABILITY: THEORY
Blazar emission varies across all wavelengths on timescales from minutes to years \citep{Rieger_microvariabilityreview_2019, raiteri_blazarvarreview_2025, andruchow_review_HBLvariabilityOptical_galaxies14030054}, and variability studies can provide constraints complementary to SED fitting.
In one-zone leptonic models, variations of the low- and high-energy components should be simultaneous in the EC case since they arise from the same electron population, or delayed by $\sim r/ (c \, \delta)$ at most in the SSC case due to the size $r$ of the emission region \citep[c is the speed of light, $\delta$ the Doppler factor, see][]{maraschi_SSC_1992ApJ...397L...5M}.
The correlation slope depends on the variability driver: electron density changes yield a quadratic relationship in SSC and a linear one in EC, whereas magnetic field variations produce a linear correlation in SSC but leave EC emission unaffected, potentially generating flares only in the low-energy band \citep{dermer_basicequations_1997ApJS..109..103D, bottcher_BLLACs_2002ApJ...581..127B}.
Single-band events are usually referred to as `orphan' flares \citep[e.g. ][]{Liodakis_CorrelationsFlaresOrphans_2019ApJ...880...32L}.
Multi-zone models introduce delays or additional orphan events; for example, mirror models \citep{Vittorini_flareBLlac_2009ApJ, vittorini_blobmodel_2014, tavani_flaremodel_2015, vittorini_3C279_model_2017, bottcher_MirrorModels_2021Physi...3.1112B} predict that flare photons photo-ionize nearby BLR clouds or other plasmoids in the jet, creating delayed EC seed photons and causing $\gamma$-rays to decouple from the optical emission.
Orphan $\gamma$-ray flares can arise in several others leptonic multi-zone setups via acceleration of plasma blobs \citep{bihan_zech_Orphan3C279_2026JHEAp..5300629L}, blob propagation through synchrotron-emitting ring of electrons \citep[ring-of-fire, see][]{MacDonald_RingofFire_2015ApJ...804..111M, Traianou_RingOfFire_3C120_2026A&A...710A.123T}, synchrotron suppression \citep{sobacchi_orphanflare_2021MNRAS.503..688S}, or magnetic reconnection \citep{zhang_simulations_reconnection_2022ApJ...924...90Z}.
Alternatively, they could also indicate hadronic processes \citep{bottcher_hadronic_orphan_2005ApJ...621..176B, cerruti_jetprocesses_2020}.

% 3 - BLAZAR VARIABILITY: STUDIES
Monitoring campaigns generally support leptonic frameworks, frequently detecting correlations between optical and $\gamma$-ray bands with time lags below $\sim 10$ days \citep{cohen_blazar_correlations_2014, majumder_mwl_variability_2019, acharyya_blazar_correlation_2023}, implying co-spatial emission.
Orphan flares are conventionally viewed as rare exceptions to one-zone leptonic models requiring multi-zone or hadronic interpretations, though recent findings indicate they are more frequent than previously assumed \citep{Liodakis_CorrelationsFlaresOrphans_2019ApJ...880...32L, dejaeger_blazarflare_correlation_2023}.
Subclass differences show that FSRQs are typically more Compton-dominated, display larger variability amplitudes, and exhibit weaker $\gamma$-ray/optical correlations than BL Lacs \citep{abdo_FermiVariability_2010ApJ...722..520A, hovatta_blazzar_correlation_2014}; this aligns with EC dominance in low-synchrotron-peak objects and SSC in high-synchrotron-peak BL Lacs \citep{finke_CD_blazarsequence_2013ApJ...763..134F}.

% 4 - FAST VARIABILITY
Blazars also exhibit fast variability on timescales of hours or minutes \citep{foschini_FSRQ_shortvariaiblitysearch_2011, Saito_PKS1510_variablity_2013, Liao_PKS0507_2015, ackermann_3C279minutes_2016ApJ...824L..20A, gupta_blazar_qpo_idv_2018, shukla_nature_3C279minutes_2020, raiteri_S4_0954_extremevariability_2023, LST1_BLLACVariability_2026A&A...710A..41A}.
By causality, the variability timescale restricts the co-moving size of the emitting region \citep{Rieger_microvariabilityreview_2019}.
While high Doppler factors can relax this constraint, the values required for the fastest flares are often unphysical, pointing instead to compact jet substructures rather than a single extended zone \citep{aharonian_ultrafastvariability_2017ApJ...841...61A}.
Upper limits on black hole masses inferred from these short timescales are typically lower than estimates from host-galaxy scaling relations, further indicating that fast flares require complex sub-jet geometries or localized acceleration zones \citep{salvati_mkn421_variability_1998, giannios_jets_in_jet_2009, Rieger_microvariabilityreview_2019}.
Alternatively, interpreting the variability timescale as the radiative cooling time constrains the electron energy distribution and local photon density \citep{begelman_implications_2008MNRAS.384L..19B, Liao_review_2018}.
Proposed physical mechanisms to explain fast variability include relativistic sub-regions moving within the bulk flow \citep[jets-in-jet;][]{giannios_jets_in_jet_2009}, jet-star interactions producing mini-jets \citep{barkov_jetstar_2012ApJ...749..119B, blandford_magnetoluminescence_2017SSRv..207..291B}, and particle acceleration in magnetospheric gaps near the black hole \citep{levinson_flaresFromSMBHgaps_2011ApJ...730..123L}.

% 5 - GOAL OF THE WORK
Existing blazar variability studies are subject to limitations that leave gaps in our understanding.
Most studies focus on a single band in isolation \citep[e.g. ][]{meyer_fermilat_variability_2019, fan_3C454.3_optical_2021ApJS..253...10F, dogra_3C454_variability_optical_2025}, or restrict multi-wavelength SED modelling to isolated flares \citep[e.g.][]{abdo_3c454_fermi_flare2010_2011, MAGIC_Mkn421_2015A&A...576A.126A}, preventing comparisons across different epochs or between active and quiescent states.
By contrast, long-baseline statistical studies usually employ temporal bins of one day or more \citep{cohen_blazar_correlations_2014, majumder_mwl_variability_2019, rajput_3C454_temporalcorrelation_2019, nalewajko_correlation3C454.3_2019A&A...631A...4N}, smoothing over intra-day variations despite growing evidence that fast blazar variability can be physically distinct from longer-term trends processes \citep{villata_BLLAC_2002A&A...390..407V, begelman_implications_2008MNRAS.384L..19B, aharonian_ultrafastvariability_2017ApJ...841...61A, raiteri_BLLAC_maximum_2023MNRAS.522..102R, raiteri_blazarvarreview_2025}.
Datasets combining multi-band coverage, long multi-epoch baselines, and intra-day resolution remain rare.
The bright FSRQ 3C~454.3 is an ideal target for such analysis, as its high flux allows fine temporal binning across most of its dynamic range.

% 6 - CONCLUSION
Here we present a systematic optical and $\gamma$-ray correlation study of 3C~454.3 at intra-day resolution over a $17$-year baseline, comparing multiple flaring and quiescent epochs.
The paper is structured as follows: Sect.~\ref{sec:3C454.3} reviews the known correlation properties of 3C~454.3; Sect.s~\ref{sec:data:optical} and~\ref{sec:data_analysis} describe the optical and $\gamma$-ray observations and analysis pipelines; Sect.~\ref{sec:global_lightcurve} examines the full dataset correlation; Sect.s~\ref{sec:flares} and~\ref{sec:spectra} detail the temporal and spectral behaviour of individual flaring epochs; and Sect.s~\ref{sec:discussion} and~\ref{sec:conclusions} provide the discussion and summary.

%%%%%%%%%%%%%%%%%%%%%%%%%%%%%%%%%%%%%%%%%%
%%%%%%%%%%%%%%%%%%%%%%%%%%%%%%%%%%%%%%%%%%
%%%%%%%%%%%%%%%%%%%%%%%%%%%%%%%%%%%%%%%%%%
\section{3C 454.3: variability and optical-\texorpdfstring{$\gamma$}{gamma}-ray correlation}
\label{sec:3C454.3}
% 1 - SOURCE PROPERTIES
3C~454.3 is a highly variable FSRQ at redshift $z=0.859$ \citep{Jackson_redshiftmeasure_1991}, hosting a supermassive black hole of mass $M_\bullet= 10^{8.83} \si{\solarmass}$ estimated through virial assumption using the H$\beta$ line-width and BLR
size determined from reverberation mapping \citep{woo+urry_2002_BHmass_ApJ...579..530W, Gu_2001_BHMASS_MNRAS.327.1111G}.
As one of the brightest and most studied $\gamma$-ray sources in the sky, it provides an ideal laboratory for multi-epoch, fast variability studies.
Its small jet viewing angle of $\lesssim \SI{1}{\degree}-\SI{2}{\degree}$ \citep{jorstad_viewingangles_2017ApJ...846...98J} yields strong Doppler boosting that amplifies both the observed flux and the variability amplitude, making fast signatures more readily detectable than in less favourably oriented sources.

% 2 - SINGLE FLARE STUDIES
The source has exhibited several major multi-wavelength outbursts.
Notable activity periods occurred in 2005–2006 \citep{villata_3c454.3flaresWEBT_2006A&A...453..817V, giommi_3C454.3Flare_2006A&A...456..911G, pian_3C454.3Flare05Integral_2006A&A...449L..21P, jorstad_3C454.3Flares0508_2010ApJ...715..362J} and 2007–2010 \citep{ghisellini_3C454.3Flare07_AGILE_2007MNRAS.382L..82G, raiteri_3C454.3Flare07_Flare2008A&A...491..755R, abdo_3C454.3Flare2008_2009ApJ...699..817A, vercellone_3C454.3Flare07AGILE_2009ApJ...690.1018V, Donnarumma_3C454.3Flare07AGILE_2009ApJ...707.1115D, vercellone_3C454.3FlaresPointingAGILE_2010ApJ...712..405V}, culminating in an exceptional $\gamma$-ray flare in November 2010 \citep{Vercellone_3C454flare2010_2011, abdo_3c454_fermi_flare2010_2011, raiteri_3C454_optical_2011, wehrle_3C454.3Flare10Conic_2012ApJ...758...72W, jorstad_3C454.3Flares0910_2013ApJ...773..147J}, followed by renewed activity in 2014–2016 \citep{britto_3C454.3Flare14_2016ApJ...830..162B, coogan_3C454.3Flare14_2016MNRAS.458..354C, Kushwaha_3C454.3Flare14_2017MNRAS.464.2046K, weaver_3C454_june2016_2019, Liodakis_3C454.3Flare1314_2020ApJ...902...61L}.
Broadband SED modelling of these individual flares has produced diverse interpretations, variously attributing the high-energy emission to one-zone EC, SSC, mixed EC+SSC, or multi-zone EC scenarios.
This variety underscores the limitations of single-flare investigations and reinforces the need for a systematic, long-term approach to identify consistent trends across epochs.

% 3 - CORRELATION STUDIES
On timescales exceeding one day, statistical correlation studies consistently show a significant optical–$\gamma$-ray correlation with no systematic time lags \citep{Bonning_3C_opt_gamma_correlation_2009, vercellone_3C454.3FlaresPointingAGILE_2010ApJ...712..405V, Kushwaha_3C454.3Flare14_2017MNRAS.464.2046K}; typical discrete correlation function values peak between $0.4$ and $0.7$.
The flux relationship follows a power law, $F_\gamma \propto F_{\rm opt}^\eta$, with an index of $\eta \approx 1-2$ for R-band fluxes \citep{bonnoli_3C454.3Correlation09_2011MNRAS.410..368B, chatterjee_Similarity_2012ApJ...749..191C} and a shallower $\eta \approx 0.6-1.5$ for the V-band \citep{rajput_3C454_temporalcorrelation_2019}.
These values are broadly consistent with variability driven by fluctuations in the magnetic field, electron density, or Doppler factor within leptonic frameworks.
Additionally, correlations between long-integration $\gamma$-ray and radio core emission (at $15$ and $\SI{43}{\giga\hertz}$) have been reported, with results favouring SSC as the dominant $\gamma$-ray emission mechanism \citep{Palafox_3C454.3RadioGamma_2025A&A...696A..70P}.

% 4 - FAST VARIABILITY
At shorter timescales, fast variability has been detected during specific epochs, including $2-5 \text{ h}$ variations in December 2009 \citep{tavecchio_RapidVariability_2010MNRAS.405L..94T, raiteri_3C454_optical_2011}, April 2010 \citep{foschini_3C454.3FlareApr10_2010MNRAS.408..448F}, and June 2016 \citep{weaver_3C454_june2016_2019}, alongside sub-hour fluctuations in May–June 2014 \citep{britto_3C454.3Flare14_2016ApJ...830..162B, coogan_3C454.3Flare14_2016MNRAS.458..354C}.
However, these analyses remain restricted to isolated events.
By combining data from multiple flaring and quiescent epochs systematically, this work aims to establish whether fast variability is a persistent characteristic across the full dynamic range of the source, bridging the gap between detailed single-flare modelling and long-term statistical studies.

%%%%%%%%%%%%%%%%%%%%%%%%%%%%%%%%%%%%%%%%%%
\begin{figure*}
   \centering
   \resizebox{\hsize}{!}{\includegraphics{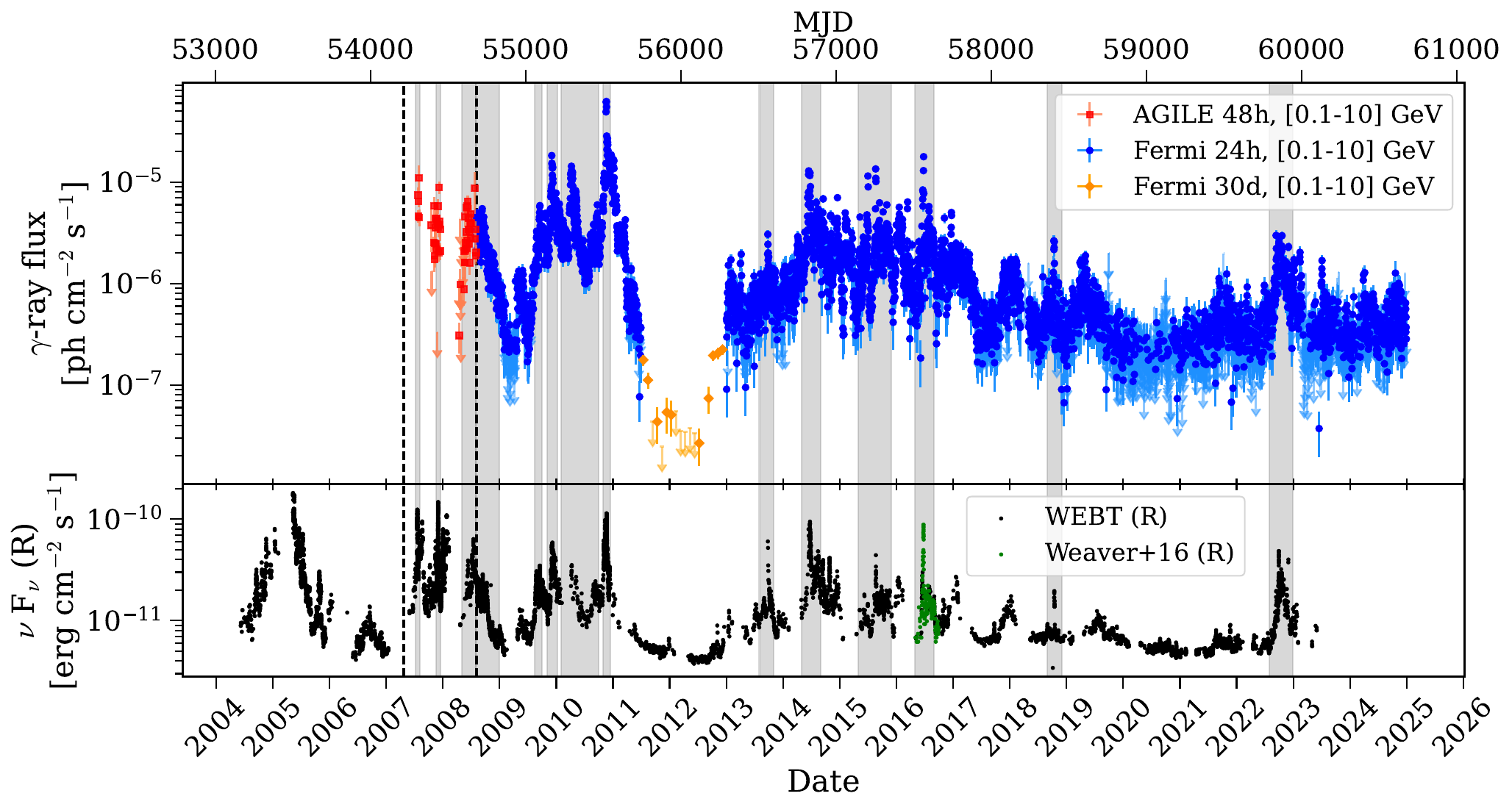}}
   \caption{$\gamma$-ray and optical $R$-band light curves of 3C~454.3. Vertical dashed lines indicate the start of AGILE (MJD 54214) and \textit{Fermi} (MJD 54683) monitoring. Top panel: $0.1-10~\si{\GeV}$ light curves. Red squares show the $\SI{48}{\hour}$ AGILE data during the AGILE-only period; blue circles the $\SI{24}{\hour}$ \textit{Fermi} light curve; orange diamonds indicate the $\SI{30}{\day}$ binned \textit{Fermi} flux during the 2011-2013 quiescent period (MJD 55743-56293); arrows represent $95\%$ confidence level upper limits. Bottom panel: $R$-band optical light curve. Grey vertical shaded areas mark the individual flaring epochs selected in Table~\ref{table:Periods}.
   }
 \label{fig:LightCurve_overview}
\end{figure*}
%%%%%%%%%%%%%%%%%%%%%%%%%%%%%%%%%%%%%%%%%%

%%%%%%%%%%%%%%%%%%%%%%%%%%%%%%%%%%%%%%%%%%
%%%%%%%%%%%%%%%%%%%%%%%%%%%%%%%%%%%%%%%%%%
%%%%%%%%%%%%%%%%%%%%%%%%%%%%%%%%%%%%%%%%%%
\section{The optical data sample}
\label{sec:data:optical}
% THE DATASET
Most of the optical data used in this work were provided by the Whole Earth Blazar Telescope \citep[WEBT;][]{villata_3c454.3flaresWEBT_2006A&A...453..817V, villata_2007A&A...464L...5V, villata_2009A&A...504L...9V, raiteri+_2007A&A...473..819R, raiteri+_2008A&A...485L..17R, raiteri_3C454.3Flare07_Flare2008A&A...491..755R, raiteri_3C454_optical_2011, dogra_3C454_variability_optical_2025},\footnote{\url{https://www.oato.inaf.it/blazars/webt/}} an international collaboration of optical, near-infrared, and radio observatories that monitors $\gamma$-ray-loud blazars with high temporal density.
A comprehensive description of the data reduction, cross-calibration, and merging procedures for this dataset is provided by \citet{dogra_3C454_variability_optical_2025}.
The baseline spans from June 2004 to June 2023, incorporating archival WEBT observation campaigns for 3C~454.3 alongside public observatory data from SMARTS \citep{bonning_SMARTS_2012ApJ...756...13B} and the Zwicky Transient Facility \citep[ZTF;][]{Bellm_ZTF_2019PASP..131a8002B}.
The observations were carried out using CCD detectors equipped with broadband Johnson-Cousins filters \citep{Bessel_magnitudes_1998}.
We restrict our analysis to the $R$-band data because they constitute the largest and most densely sampled subset.
Magnitudes were converted to linear fluxes assuming the standard zero-magnitude flux of the Vega system, $f_{\nu\text{,R}} = \SI{3.064E-20}{\erg\per\second\per\square\centi\meter\per\hertz}$ at an effective wavelength\footnote{Corresponding to an effective frequency $\nu_\text{eff,R} = \SI{4.68E14}{\hertz}$ and photon energy $E_\text{eff,R} = \SI{1.93}{\electronvolt}$.} of $\lambda_\text{eff,R} = \SI{0.641}{\micro\meter}$ \citep{Bessel_magnitudes_1998}.
Additionally, we supplemented this dataset with the $R$-band fluxes from \citet{weaver_3C454_june2016_2019} to ensure continuous coverage during the major June 2016 flare.
The resulting long-term optical light curve is displayed in the bottom panel of Fig.~\ref{fig:LightCurve_overview}.

% COMPONENTS OF THE OPTICAL EMISSION
The $R$-band emission of 3C~454.3 is governed by two distinct physical components: non-thermal synchrotron radiation originating from the relativistic jet (which typically peaks in the infrared regime) and thermal emission from the accretion disc (which peaks in the near-ultraviolet and produces the `big blue bump', BBB, in the broadband SED; \citealt{malkan_BBB_1982ApJ...254...22M, raiteri+_2007A&A...473..819R, raiteri_3C454_optical_2011}).
While the jet synchrotron component remains present even during the quiescent states \citep{dogra_3C454_variability_optical_2025}, the relative fractional contribution of these two components shifts dramatically with overall activity.
The thermal disc emission becomes prominent during low-flux states, whereas the non-thermal jet emission completely overwhelms the disc during high-activity phases \citep{Amaya_States3C454.3_2021ApJ}.
This induces an overall `redder-when-brighter' (RWB) spectral trend across the optical bands \citep{villata_3c454.3flaresWEBT_2006A&A...453..817V, Bonning_3C_opt_gamma_correlation_2009, safna_RWBtrends_2020MNRAS}.
However, the WEBT dataset highlights a more nuanced behaviour: a steep RWB trend persists up to magnitude $m_I \approx 14.5$ across several colour indices, beyond which the jet component heavily dominates, causing the trend to flatten or become nearly achromatic \citep{villata_3c454.3flaresWEBT_2006A&A...453..817V}.
This transition between disc-prominent and jet-dominated regimes is relevant for interpreting the optical–$\gamma$-ray correlation, as the thermal disc component does not contribute to the $\gamma$-ray emission and its underlying flux can dilute the observed correlation signal at low flux levels.

%%%%%%%%%%%%%%%%%%%%%%%%%%%%%%%%%%%%%%%%%%
%%%%%%%%%%%%%%%%%%%%%%%%%%%%%%%%%%%%%%%%%%
%%%%%%%%%%%%%%%%%%%%%%%%%%%%%%%%%%%%%%%%%%
\section{Gamma-ray observations and data analysis}
\label{sec:data_analysis}
The $\gamma$-ray analysis uses $0.1-10 \,\si{\giga\electronvolt}$ data from Fermi-LAT \citep{Atwood_FERMIsatellite_2009} covering the period from 2008~August~5 to 2025~January~1 (MJD 54683-60676), and from AGILE-GRID \citep{Tavani_AGILE_2008NIMPA.588...52T, Tavani_AGILE_2009A&A...502..995T} covering the period from 2007~April~24 to 2024~January~18 (MJD 54214-60327).

%%%%%%%%%%%%%%%%%%%%%%%%%%%%%%%%%%%%%%%%%%
%%%%%%%%%%%%%%%%%%%%%%%%%%%%%%%%%%%%%%%%%%
%%%%%%%%%%%%%%%%%%%%%%%%%%%%%%%%%%%%%%%%%%
\subsection{Fermi-LAT observations and data analysis}
\label{sec:data_analysis:sub:Fermi}
% FERMI SETUP
Fermi-LAT data were analysed using \texttt{fermipy 1.4.0} and Fermi Science Tools \texttt{v2.4.9} within the \texttt{FermiBottle}\footnote{\url{https://hub.docker.com/r/fssc/fermibottle/}} \texttt{docker} container.
We selected \texttt{Pass 8} photon and spacecraft data, the \texttt{iso\_P8R3\_SOURCE\_V3\_v1.txt} isotropic template, the \texttt{gll\_iem\_v07.fits} Galactic diffuse emission model, and \texttt{evclass=128}.
Standard quality cuts were applied: \texttt{evtype=3}, \texttt{zmax=90} to avoid Earth limb contamination, \texttt{DATA\_QUAL>0}, and \texttt{LAT\_CONFIG=1}.
The initial source model was built from the 4FGL-DR4 catalogue \citep{Fermi_4FGL_2020, Ballet_4FGL_DR4_2023}.
The Region of Interest (ROI) was centred on 3C~454.3 (4FGL~J2253.9$+$1609) at $(\text{R.A.},\text{ Dec})=(343.496, 16.151)\,\si{\deg}$ with a $\SI{10}{\deg}$ radius, using a binning of $\SI{0.1}{\deg}$ per pixel and eight energy bins per decade.
Flux light curves were extracted in the $0.1-10 \,\si{\giga\electronvolt}$ range at pre-defined $\SI{24}{\hour}$, $\SI{12}{\hour}$, and $\SI{6}{\hour}$ binning via maximum likelihood estimation (MLE).

% BACKGROUND
First, the three background emission coefficients (isotropic diffuse normalisation $g_\text{iso}$, Galactic diffuse normalisation $g_\text{gal}$, and Galactic spectral index $i_\text{gal}$) were determined monthly.
Using the \texttt{fermipy} dedicated methods, data selection, binning, and model generation were executed, followed by preliminary ROI parameter optimisation.
We then freed the background parameters and the normalisations of sources within $\SI{5}{\deg}$ of the ROI centre for an initial fit.
Sources with TS $<1$ or non-converged TS calculations were removed before performing a second fit freeing only the background coefficients.
The resulting stable distributions yielded average values and standard deviations of $g_\text{iso}=1.1\pm0.5$, $g_\text{gal}=0.90\pm0.21$, and $i_\text{gal}=-0.03\pm0.12$.

% LIGHT CURVE
The monthly background coefficients were fixed for the light-curve calculation, which adopted a similar two-step fitting procedure.
Following ROI setup and optimisation, the normalisation and spectral parameters of 3C~454.3 and the normalisations of sources within $\SI{5}{\deg}$ were freed for a first fit.
Sources with TS $<1$ or non-converging TS were removed, and a second fit was performed freeing only 3C~454.3.
The final light-curve fluxes for the $\SI{24}{\hour}$, $\SI{12}{\hour}$, and $\SI{6}{\hour}$ bins were then calculated by freeing the parameters of 3C~454.3 and the normalisations of sources within $\SI{5}{\deg}$ to track fast variability.
In all fits, the spectral model of 3C~454.3 is described by a log-parabola model:
\begin{equation}
    \label{eq:LP}
    \frac{\d N}{\d E} (E) = N_0 \left(\frac{E}{E_b}\right)^{-\left(\alpha+\beta \ln \left(E/E_b\right)\right)},
\end{equation}
where $N_0$ is the normalisation in $\si{\photon\per\second\per\centi\meter\squared\per\mega\electronvolt}$, $E_b=\SI{908}{\mega\electronvolt}$ the fixed scale energy, $\alpha$ the power law index, $\beta$ the curvature parameter.

% DATA QUALITY
We computed $95\%$ confidence level upper limits for bins where the test statistic \citep[TS;][]{Mattox_Likelihood_EGRET_1996} was $<9$.
We removed $\approx 8\%$ of the data points due to low quality, defined as non-converged fits (\texttt{fit\_status}$\neq0$ or \texttt{fit\_success}$=$\texttt{False}), large relative errors ($>50\%$) from low exposure, or upper limits exceeding $\SI{2E-6}{\photon\per\second\per\centi\meter\squared}$.

%%%%%%%%%%%%%%%%%%%%%%%%%%%%%%%%%%%%%%%%%%
\subsection{AGILE-GRID observations and data analysis}
\label{sec:data_analysis:sub:AGILE}
% AGILE SETUP
The AGILE spacecraft operated in `pointing mode' from the beginning of the mission until 15 October 2009 (MJD 55119), completing $101$ observation blocks of predefined long exposures that drifted by $\approx\SI{1}{\deg}$ per day due to solar panel constraints.
In November 2009, scientific operations transitioned to `spinning mode', scanning $\approx 80\%$ of the sky daily (exposure $\approx \SI{7E6}{\centi\meter\squared\second}$) at an angular velocity of $\approx\SI{0.8}{\deg\per\second}$, performing $200$ passes per day over the same region.
Mission operations are detailed in \citet{Barbiellini_AGILE_ST_2001AIPC..587..754B, Prest_AGILE_ST_2003NIMPA.501..280P, Perotti_AGILE_ACS_2006NIMPA.556..228P, Feroci_AGILE_SA_2007NIMPA.581..728F, Tavani_AGILE_2008NIMPA.588...52T, Tavani_AGILE_2009A&A...502..995T, Bulgarelli_AGILE_tracker_2010NIMPA.614..213B, Cattaneo_AGILE_Calib_2011NIMPA.630..251C, Chen_AGILE_calib_2013A&A...558A..37C, Bulgarelli_AGILE_ALERTS_2014ApJ...781...19B, Pittori_AGILE_DataCenter_2019RLSFN..30S.217P}, while the AGILE/GRID instrument, response functions, and analysis methods are described in \citet{bulgarelli_MLE_2012A&A...540A..79B, bulgarelli_2AGL_2019A&A...627A..13B, bulgarelli_agilepy_2022}.
%\footnote{\url{https://agile.ssdc.asi.it/current_pointing.html}}

% AGILE SOFTWARE
Data analysis was performed using \texttt{agilepy v1.7.0}\footnote{\url{https://agilepy.readthedocs.io}} \citep{bulgarelli_agilepy1.7.0_2025zndo} with the \texttt{FM3.119} on-ground background event filter, instrument response functions \texttt{H0025}, and Science Tools \texttt{BUILD26}.
We selected confirmed $\gamma$-ray events (\texttt{G} class) after applying standard South Atlantic Anomaly cuts and an $\SI{80}{\degree}$ Earth albedo filter.
Counts, exposure, and background maps were generated in the $0.1-10 \,\si{\giga\electronvolt}$ range over a $\SI{30}{\deg}$ region centred on 3C~454.3, with a bin size of $\SI{0.25}{\deg}$.

% ANALYSIS STRATEGY
The AGILE analysis employs the same two-step fitting approach used for Fermi.
Good time intervals (GTIs) were determined via aperture photometry centred on 3C~454.3 using a $\SI{2}{\deg}$ radius and $\SI{24}{\hour}$ binning.
A GTI is defined as a period of at least $\SI{48}{\hour}$ with daily exposure $E_\text{24h}\geq \SI{3.024E6}{\centi\meter\squared\second}$ and daily counts $N_\text{24h}\geq 5$.
The Galactic background normalisation was fixed at $a_\text{gal}=0.7$, while the isotropic background normalisation $a_\text{iso}$ was estimated on a $14$-day basis using only data within the GTIs, yielding a mean and standard deviation of $a_\text{iso}=8.4 \pm 1.2$.
These background values were fixed for the light-curve calculation, which modelled 3C~454.3 using Eq.~\eqref{eq:LP}.
Due to AGILE's lower effective area and photon statistics compared to Fermi, the spectral parameters $\alpha$ and $\beta$ were estimated on a $\SI{48}{\hour}$ basis (with two bins per decade) and kept fixed during the flux estimation of the shorter $\SI{24}{\hour}$, $\SI{12}{\hour}$, and $\SI{6}{\hour}$ bins.

%%%%%%%%%%%%%%%%%%%%%%%%%%%%%%%%%%%%%%%%%%%%%%%%%%%%%%%%%%%%%%%%%%%%%
%%%%%%%%%%%%%%%%%%%%%%%%%%%%%%%%%%%%%%%%%%%%%%%%%%%%%%%%%%%%%%%%%%%%%
%%%%%%%%%%%%%%%%%%%%%%%%%%%%%%%%%%%%%%%%%%%%%%%%%%%%%%%%%%%%%%%%%%%%%
%%%%%%%%%%%%%%%%%%%%%%%%%%%%%%%%%%%%%%%%%%%%%%%%%%%%%%%%%%%%%%%%%%%%%
%%%%%%%%%%%%%%%%%%%%%%%%%%%%%%%%%%%%%%%%%%%%%%%%%%%%%%%%%%%%%%%%%%%%%
%%%%%%%%%%%%%%%%%%%%%%%%%%%%%%%%%%%%%%%%%%%%%%%%%%%%%%%%%%%%%%%%%%%%%
%%%%%%%%%%%%%%%%%%%%%%%%%%%%%%%%%%%%%%%%%%%%%%%%%%%%%%%%%%%%%%%%%%%%%
%%%%%%%%%%%%%%%%%%%%%%%%%%%%%%%%%%%%%%%%%%%%%%%%%%%%%%%%%%%%%%%%%%%%%
%%%%%%%%%%%%%%%%%%%%%%%%%%%%%%%%%%%%%%%%%%%%%%%%%%%%%%%%%%%%%%%%%%%%%
% GLOBAL PROPERTIES

\section{Global light curve properties}
\label{sec:global_lightcurve}
% GLOBAL LIGHT CURVE
Figure~\ref{fig:LightCurve_overview} displays the $\SI{24}{\hour}$ $\gamma$-ray light curve in the $0.1-10 \,\si{\giga\electronvolt}$ range (top panel), with pre-Fermi AGILE pointing data binned at $\SI{48}{\hour}$, alongside the WEBT optical light curve (bottom panel; \citealt{dogra_3C454_variability_optical_2025}).
The source alternates between extended low and high states.
Flux declined in early 2009, followed by intense activity from mid-2009 to late 2010, including the brightest flares in the sample with peak fluxes $>\SI{E-5}{\photon\per\second\per\centi\meter\squared}$ (late 2009, early 2010, and late 2010).
An extended low state occurred from 2011 to early 2013, during which the flux fell below the weekly Fermi detection threshold for approximately 1.5 years (plotted in $\SI{30}{\day}$ bins for MJD 55743-56293). Renewed activity from 2013 to 2017 produced major flares in 2014, 2015, and 2016 ($>\SI{E-5}{\photon\per\second\per\centi\meter\squared}$), followed by a lower state with sporadic outbursts (e.g., late 2018 and 2022, $>\SI{E-6}{\photon\per\second\per\centi\meter\squared}$).
Assuming a steady baseline, \citet{meyer_fermilat_variability_2019} derived an upper limit on the quiescent flux above $\SI{100}{\mega\electronvolt}$ of $F_\text{Q}=\SI{0.86E-6}{\photon\per\second\per\centi\meter\squared}$.

%%%%%%%%%%%%%%%%%%%%%%%%%%%%%%%%%%%%%%%%%%
\begin{figure}
   \centering
   \resizebox{\hsize}{!}{\includegraphics{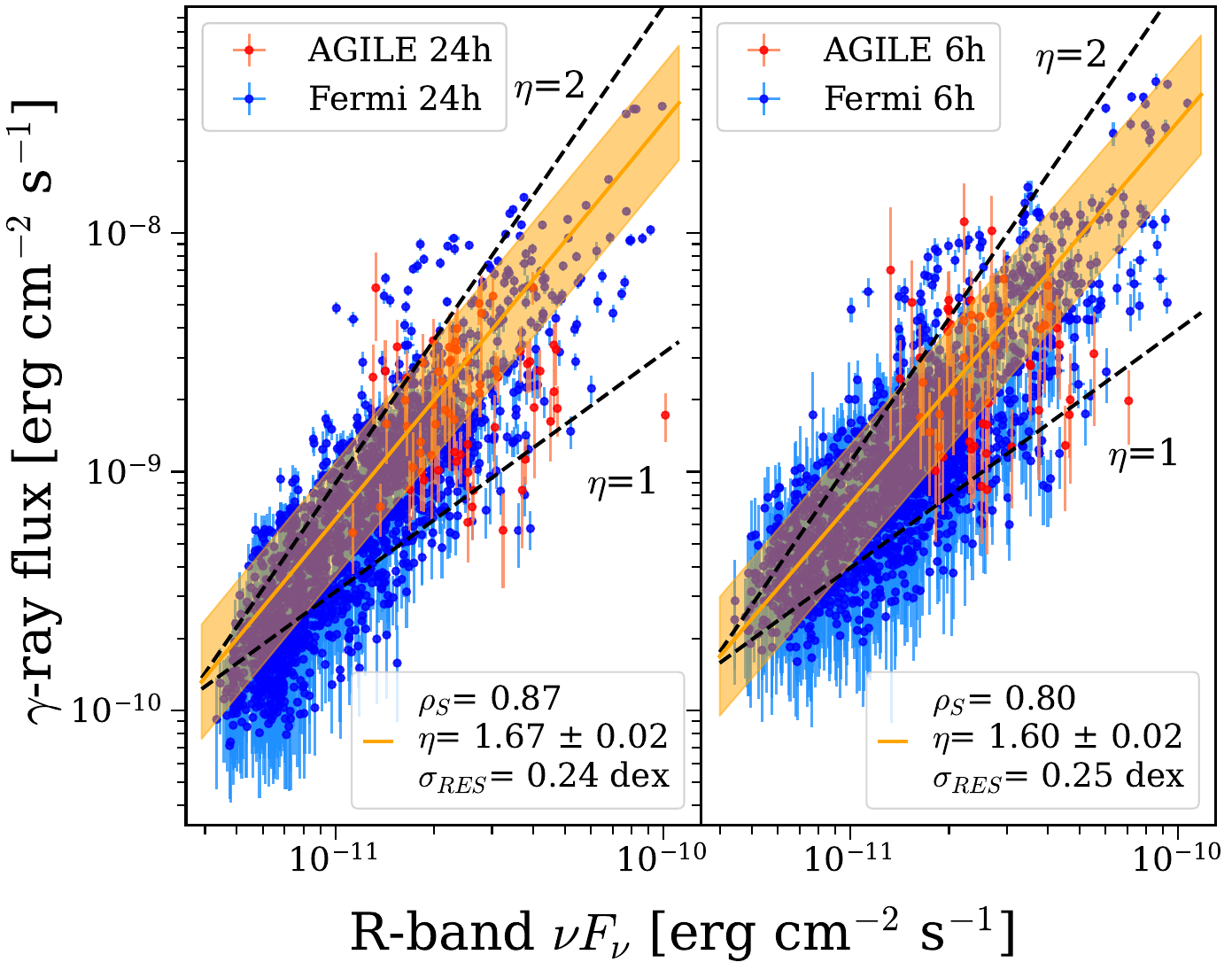}}
   \caption{
     $\gamma$-ray/optical flux correlation for the full 3C~454.3 dataset, comparing $\SI{24}{\hour}$ (left) and $\SI{6}{\hour}$ (right) temporal binning. AGILE-only data (red points) are included to extend \textit{Fermi}-LAT (blue). A strong correlation is observed with a super-linear slope, with a lower residual dispersion at the $\SI{24}{\hour}$ timescale.
   }
 \label{fig:Correlation_overview}
\end{figure}
%%%%%%%%%%%%%%%%%%%%%%%%%%%%%%%%%%%%%%%%%%

% CORRELATION
Figure~\ref{fig:Correlation_overview} presents the optical versus $\gamma$-ray energy flux correlation, excluding $\gamma$-ray upper limits, with optical data re-binned to match the $\SI{24}{\hour}$ and $\SI{6}{\hour}$ resolutions.
A strong correlation is present at both timescales, yielding Spearman coefficients of $\rho_S = 0.80$ ($\SI{6}{\hour}$) and $\rho_S = 0.87$ ($\SI{24}{\hour}$).
Assuming $L_\gamma \propto L_R^\eta$, the best-fit slopes are $\eta = 1.67\pm0.02$ ($\SI{6}{\hour}$) and $\eta = 1.60\pm0.02$ ($\SI{24}{\hour}$).
This $\sim4\%$ discrepancy likely arises because the $\SI{6}{\hour}$ light curve lacks the low-flux detections that require longer exposures.
This correlation is consistent with previous sub-sample analyses \citep{Bonning_3C_opt_gamma_correlation_2009, bonnoli_3C454.3Correlation09_2011MNRAS.410..368B} and supports a leptonic synchrotron and inverse-Compton origin.
However, the slope $\eta$ is incompatible with canonical one-zone models driven by electron density variations, which predict $\eta=1$ for EC and $\eta=2$ for SSC emission \citep{dermer_basicequations_1997ApJS..109..103D}.
This indicates either mixed EC and SSC processes or multi-zone emission.
Alternatively, a single global fit may obscure independent correlation regimes associated with different source states; this is addressed via the epoch-resolved analysis in Sect.~\ref{sec:flares}.
The higher $\rho_S$ and lower logarithmic residual dispersion ($\sigma_\text{RES}$) at $\SI{24}{\hour}$ might reflect both the larger uncertainties in the $\SI{6}{\hour}$ data and the smoothing of short-term intrinsic fluctuations.

%%%%%%%%%%%%%%%%%%%%%%%%%%%%%%%%%%%%%%%%%%
\begin{figure}
   \centering
   \resizebox{\hsize}{!}{\includegraphics{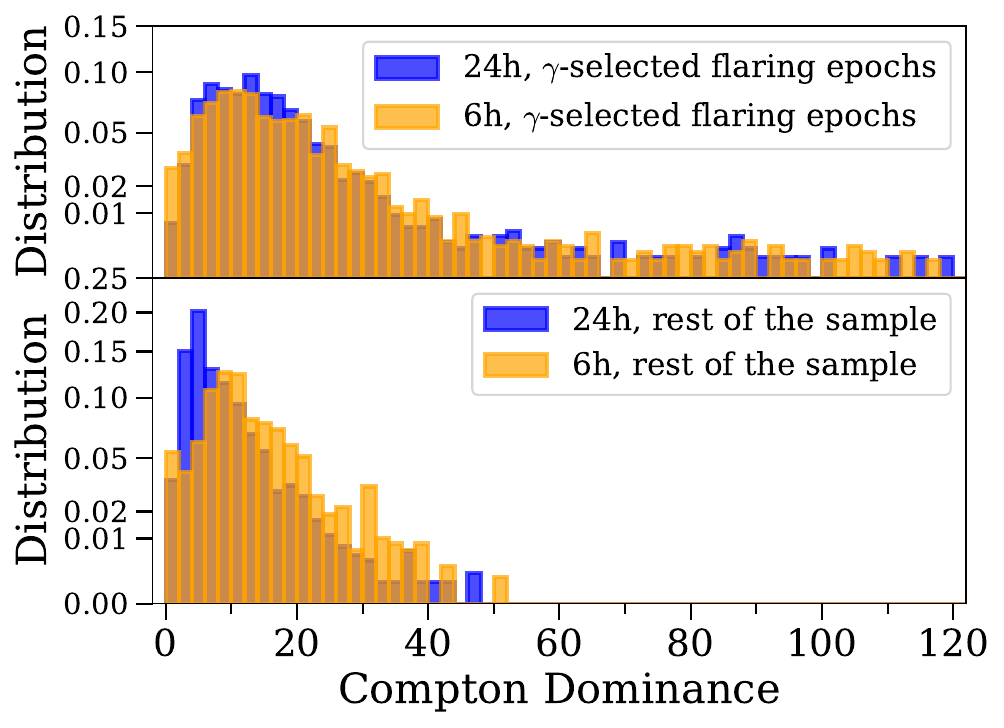}}
   \caption{Distribution of the Compton dominance during active flaring epochs (top panel; listed in Table~\ref{table:Periods}) compared to the rest of the 17-year monitoring period (bottom panel). Distributions are evaluated at $\SI{6}{\hour}$ (orange) and $\SI{24}{\hour}$ (blue) temporal resolutions. Active intervals exhibit a prominent high-$\mathrm{CD}$ tail extending well beyond baseline levels.
   The strongest difference between $\SI{6}{\hour}$ and $\SI{24}{\hour}$ is seen in the off-flare distributions, suggesting 3C~454.3 can present fast flares also outside strong flaring epochs, which are reduced in the $\SI{24}{\hour}$ histogram due to time-averaging with lower-CD emission.
   }
 \label{fig:ComptonDominance}
\end{figure}
%%%%%%%%%%%%%%%%%%%%%%%%%%%%%%%%%%%%%%%%%%

% COMPTON DOMINANCE
The Compton dominance (CD) for each temporal bin is defined as:
\begin{equation}
    \label{eq:ComptonDominance}
    \text{CD} = \frac{\nu\, F_\nu (1\,\text{GeV})}{\nu\, F_\nu (\text{R})},
\end{equation}
where $\nu F_\nu (\SI{1}{\giga\electronvolt})$ is derived from Eq.~\eqref{eq:LP}, as $\displaystyle E^2 \frac{\d N}{\d E} (E)$, using the best-fit parameters.
This quantity serves as an observational proxy to the theoretical CD (the ratio of the true SED peak luminosities), which requires detailed spectral modelling.
Figure~\ref{fig:ComptonDominance} compares the CD distributions during the $14$ flaring epochs listed in Table~\ref{table:Periods} (top) with the rest of the sample (bottom), at both $\SI{6}{\hour}$ and $\SI{24}{\hour}$.
All distributions exhibit a low-CD component peaking at $\mathrm{CD} \approx 5-10$, typical of quiescent periods where the $R$-band flux contains a significant thermal accretion disc contribution \citep{Amaya_States3C454.3_2021ApJ, dogra_3C454_variability_optical_2025}.
By contrast, the flaring periods display a broad tail extending from $\mathrm{CD} \approx 45$ to $120$, where non-thermal jet emission completely dominates the optical band.
The off-flare distributions show the most pronounced discrepancy between the $\SI{6}{\hour}$ and $\SI{24}{\hour}$ integrations, indicating the presence of rapid, intermediate-CD ($\approx 15-40$) flares even outside major flaring epochs.

%%%%%%%%%%%%%%%%%%%%%%%%%%%%%%%%%%%%%%%%%%%%%%%%%%%%%%%%%%%%%%%%%%%%%
%%%%%%%%%%%%%%%%%%%%%%%%%%%%%%%%%%%%%%%%%%%%%%%%%%%%%%%%%%%%%%%%%%%%%
%%%%%%%%%%%%%%%%%%%%%%%%%%%%%%%%%%%%%%%%%%%%%%%%%%%%%%%%%%%%%%%%%%%%%
%%%%%%%%%%%%%%%%%%%%%%%%%%%%%%%%%%%%%%%%%%%%%%%%%%%%%%%%%%%%%%%%%%%%%
%%%%%%%%%%%%%%%%%%%%%%%%%%%%%%%%%%%%%%%%%%%%%%%%%%%%%%%%%%%%%%%%%%%%%
%%%%%%%%%%%%%%%%%%%%%%%%%%%%%%%%%%%%%%%%%%%%%%%%%%%%%%%%%%%%%%%%%%%%%
%%%%%%%%%%%%%%%%%%%%%%%%%%%%%%%%%%%%%%%%%%%%%%%%%%%%%%%%%%%%%%%%%%%%%
%%%%%%%%%%%%%%%%%%%%%%%%%%%%%%%%%%%%%%%%%%%%%%%%%%%%%%%%%%%%%%%%%%%%%
%%%%%%%%%%%%%%%%%%%%%%%%%%%%%%%%%%%%%%%%%%%%%%%%%%%%%%%%%%%%%%%%%%%%%
% FLARES

%%%%%%%%%%%%%%%%%%%%%%%%%%%%%%%%%%%%%%%%%%
\begin{table*}
 \centering
 \caption{
  Correlation properties of selected flaring epochs of 3C~454.3.
 }
 \label{table:Periods}
 \resizebox{\width}{!}{
 \begin{tabular}{lllccc}
    \hline\hline
    ID & Epoch [Date] & Epoch [MJD] & $\rho_S$ & Slope $\eta$ & $\sigma_\text{RES}$ [dex]\\
    \hline
    E07A & 2007 07/08 - 08/08 & 54289 - 54320 & & & \\
    E07B & 2007 11/19 - 12/19 & 54423 - 54453 & 0.20 & 0 .40$\pm$0.20 & 0.24 \\
    E08A & 2008 05/01 - 08/01 & 54587 - 54679 & & & \\
    \hline
    E08B & 2008 08/01 - 12/31 & 54679 - 54831 & 0.79 & 0.97$\pm$0.05 & 0.17 \\
    E09A & 2009 08/15 - 10/01 & 55058 - 55105 & 0.23 & 0.73$\pm$0.20 & 0.20 \\
    E09B & 2009 11/05 - 2010 01/10 & 55140 - 55206 & 0.81 & 1.29$\pm$0.08 & 0.18 \\
    E10A & 2010 02/01 - 10/01 & 55228 - 55470 & 0.59 & 1.86$\pm$0.15 & 0.22 \\
    E10B & 2010 10/27 - 12/15 & 55496 - 55545 & 0.36 & 1.14$\pm$0.10 & 0.24 \\
    E13  & 2013 08/01 - 11/01 & 56505 - 56597 & 0.48 & 0.99$\pm$0.10 & 0.15 \\
    E14  & 2014 05/01 - 09/01 & 56778 - 56901 & 0.73 & 1.33$\pm$0.08 & 0.18 \\
    E15  & 2015 05/01 - 12/01 & 57143 - 57357 & 0.48 & 1.61$\pm$0.13 & 0.22 \\
    E16  & 2016 05/01 - 09/01 & 57509 - 57632 & 0.64 & 1.26$\pm$0.05 & 0.18 \\
    E18  & 2018 09/01 - 12/01 & 58362 - 58453 & 0.56 & 2.00$\pm$0.20 & 0.18 \\
    E22  & 2022 08/01 - 12/31 & 59792 - 59944 & 0.63 & 0.53$\pm$0.09 & 0.17 \\
    \hline
    Full & sample & 54289 - 60676 & 0.80 & 1.60$\pm$0.02 & 0.25 \\
    \hline\hline
 \end{tabular}
 }
  \tablefoot{
  Full-sample and \textit{Fermi} results are computed using $\SI{6}{\hour}$ light curves for $\gamma$-ray and optical fluxes. The first three epochs are AGILE-only and are considered together using $\SI{24}{\hour}$ light curves. $\rho_S$ is the Spearman correlation coefficient, $\eta$ the best-fit slope of a linear fit in log-log space, and $\sigma_\text{RES}$ the standard deviation of the residuals.
  }
\end{table*}
%%%%%%%%%%%%%%%%%%%%%%%%%%%%%%%%%%%%%%%%%%

%%%%%%%%%%%%%%%%%%%%%%%%%%%%%%%%%%%%%%%%%%
\begin{figure*}
   \centering
   \resizebox{\hsize}{!}{\includegraphics{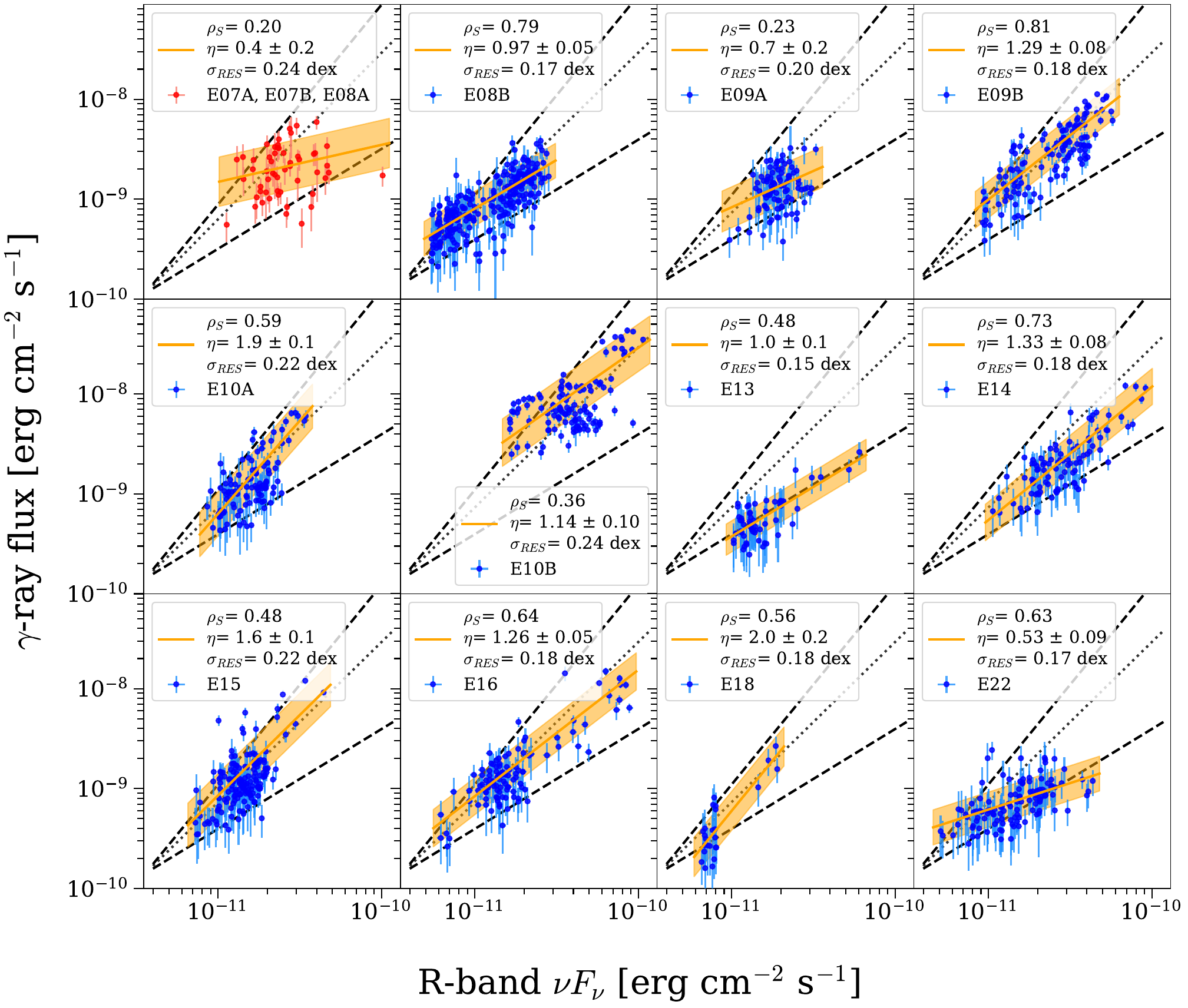}}
   \caption{Correlation of the $\gamma$-ray and optical fluxes of 3C~454.3 during each flaring epoch of Table~\ref{table:Periods} using \textit{Fermi} $\SI{6}{\hour}$ light curves (blue points) and AGILE $\SI{24}{\hour}$ light curves (red points) for the AGILE-only epochs. We report the Spearman correlation $\rho_S$, the best-fit correlation line with slope $\eta$, the dispersion $\sigma_\text{RES}$, and the $1\sigma$ uncertainty band for each period. The dotted line shows the best-fit correlation slope of the $\SI{6}{\hour}$ full sample at $\eta=1.60$; dashed lines show $\eta=1$ and $\eta=2$.}
   \label{fig:Correlation_flares}
\end{figure*}
%%%%%%%%%%%%%%%%%%%%%%%%%%%%%%%%%%%%%%%%%%

\section{Properties of selected flaring epochs}
\label{sec:flares}
We selected 14 `flaring epochs' of 3C~454.3, defined as periods of heightened $\gamma$-ray activity ($\text{flux} \gtrsim \SI{2e-6}{\centi\meter^{-2}\per\second}$) with good optical coverage (multiple measurements per night), as listed in Table~\ref{table:Periods}.
In Appendix~\ref{sec:app:flares}, Figure~\ref{fig:LightCurves_flaring_episodes} presents the $0.1-10$ GeV and $R$-band light curves for these epochs alongside the temporal evolution of the CD.
The corresponding $\gamma$-ray versus optical correlation plots are displayed in Fig.~\ref{fig:Correlation_flares} and summarised in Table~\ref{table:Periods}.
While most selected epochs exhibit a common global structure consisting of a baseline plateau with superimposed outbursts, the detailed temporal profiles vary considerably between individual periods, revealing diverse variability patterns.

%%%%%%%%%%%%%%%%%%%%%%%%%%%%%%%%%%%%%%%%%%%%%
\subsection{Optical-\texorpdfstring{$\gamma$}{gamma}-ray correlation strength and dispersion}
\label{sec:flares:sub:correlation_coefficient}
The strong correlation observed for the $\SI{6}{\hour}$ full sample ($\rho_S = 0.80$) decomposes across individual flaring epochs into a broad range of correlation strengths, spanning $\rho_S = 0.20$ to $0.81$.
Epochs E08B, E09B, and E14 are characterised by strong correlations ($\rho_S > 0.7$) and small dispersion ($\sigma_\text{RES} < 0.2~\text{dex}$), despite representing different source activity states, including a post-flare decay phase (E08B) and two high-state periods (E09B and E14).
Moderately correlated epochs (E10A, E13, E15, E16, E18, E22; $0.4 \leq \rho_S \leq 0.7$) exhibit greater diversity in variability patterns and dispersion values, likely influenced by flux uncertainties and limited sampling.
Epochs E07A, E07B, E08A, E09A, and E10B show the weakest correlations ($\rho_S < 0.4$) and the largest scatter ($\sigma_\text{RES} > 0.2~\text{dex}$), deviating significantly from a simple global trend.

Epoch E10B is the brightest and most complex flaring period in our sample, reaching the highest $\gamma$-ray flux observed from 3C~454.3.
Its optical-$\gamma$ correlation exhibits the largest dispersion of all epochs ($\sigma_\mathrm{RES}=0.24$), with data points distributed in distinct clusters rather than following a single linear trend (Fig.~\ref{fig:Correlation_flares}).
The average slope is $\eta=1.14\pm0.10$, but the large scatter indicates the coexistence of multiple emission states, likely reflecting rapidly changing physical conditions and multiple emitting regions.

Overall, these results suggest that the strong $\gamma$-ray/optical correlation typically reported for FSRQs, including 3C~454.3 \citep[e.g.][]{Bonning_3C_opt_gamma_correlation_2009, cohen_blazar_correlations_2014, majumder_mwl_variability_2019, acharyya_blazar_correlation_2023}, can break down when examined at short timescales within individual flaring epochs.
Different epochs yield substantially different Spearman coefficients and residual dispersions, with no uniform value emerging across the sample.
The strong correlation found for the full dataset thus represents an average over fundamentally distinct behaviours, indicating that the underlying variability mechanisms change between epochs and source states.

%%%%%%%%%%%%%%%%%%%%%%%%%%%%%%%%%%%%%%%%%%%%%
\subsection{Correlation slope and emission regimes}
\label{sec:flares:sub:correlation_slope}
The optical versus $\gamma$-ray flux correlation exhibits distinct scaling regimes across the flaring epochs, consistent with values from previous 3C~454.3 monitoring campaigns \citep{bonnoli_3C454.3Correlation09_2011MNRAS.410..368B, chatterjee_Similarity_2012ApJ...749..191C}.
A linear scaling regime ($\eta \sim 1$) occurs in epochs E08B and E13, matching particle-driven variability in EC-dominated one-zone scenarios.
A super-linear regime ($\eta \sim 1.2$--$1.6$) is observed in epochs E09B, E10B, E14, E15, and E16, indicating an enhanced $\gamma$-ray response driven by variations in the properties of the seed photons field or physical conditions within the jet.
By contrast, epochs E10A and E18 show a nearly quadratic scaling ($\eta \sim 2$); while in general the quadratic scaling can suggest SSC scenarios, the CD values of these epochs in particular are too high for a simple one-zone SSC model (see Sect.~\ref{sec:discussion}).
The remaining epochs (E07A, E07B, E08A, E09A, E22) yield low Spearman coefficients ($\rho_S \lesssim 0.3$), except for E22 ($\rho_S=0.63$, $\eta=0.53\pm0.09$).
The fitted slopes in these weakly correlated intervals likely reflect uncorrelated variability, orphan flares, or large observational uncertainties rather than a true sub-linear scaling regime.
The coexistence of linear, super-linear, quadratic, and uncorrelated scaling behaviours across epochs demonstrates that the $\gamma$-ray/optical correlation of 3C~454.3 cannot be explained by a single emission mechanism or a static one-zone model, but instead reflects diverse, epoch-dependent variability origins.

%%%%%%%%%%%%%%%%%%%%%%%%%%%%%%%%%%%%%%%%%%%%%
\subsection{Compton dominance evolution}
\label{sec:flares:sub:compton_dominance}
Figure~\ref{fig:LightCurves_flaring_episodes} displays diverse $\mathrm{CD}$ evolution patterns across the flaring epochs.
During $\gamma$-ray plateaus, the $\mathrm{CD}$ typically maintains moderate values ($\approx 10-30$), which exceed the global off-flare baseline (Fig.~\ref{fig:ComptonDominance}).
Individual epochs show distinct evolution patterns:
\begin{itemize}
    \item \textbf{Stable CD with fast excursions} (E08B, E14, E15, E16, E18, E22): The $\mathrm{CD}$ remains approximately constant outside of rapid outbursts. Epoch E15 is the most extreme case, repeatedly reaching $\mathrm{CD} \sim 80$ within six hours. This pattern also characterises fast orphan flares in E15 and E16, which manifest as brief $\mathrm{CD}$ spikes.
    \item \textbf{Asynchronous band evolution} (E09A, and likely E07A, E07B, E08A): $\mathrm{CD}$ changes are driven by mismatched variability rates between the optical and $\gamma$-ray bands rather than distinct correlated or orphan flares. AGILE-only epochs are less constrained due to sparser simultaneous sampling.
    \item \textbf{Gradual CD variation} (E09B, E10A): The $\mathrm{CD}$ varies slowly, driven by a super-linear correlated flux increase in both bands, such as the $\mathrm{CD}$ rise to $\sim 60$ in E09B ($\eta=1.29\pm0.08$) and the post-flare decay to $\mathrm{CD} \sim 20$ in E10A ($\eta=1.86\pm0.15$).
    \item \textbf{Constant CD during flare} (E13): The $\mathrm{CD}$ remains steady throughout the outburst due to linear correlated scaling ($\eta \simeq 1$).
    \item \textbf{Highly structured evolution} (E10B): Multiple behaviours coexist within this single complex epoch. Daily sampling suggests a gradual increase during the main flare, but intra-day resolution reveals rapid excursions between $\sim 40$ and $\sim 120$ in $\lesssim \SI{3}{\hour}$. These rapid variations indicate that the synchrotron and IC components can decouple on timescales $<\SI{24}{\hour}$, which is difficult to explain in one-zone scenarios \citep[e.g.][]{Rieger_microvariabilityreview_2019}.
\end{itemize}
Epochs with similar correlation properties can exhibit markedly different $\mathrm{CD}$ evolution, demonstrating that global correlation metrics do not uniquely constrain the energy partition between emission components.

%%%%%%%%%%%%%%%%%%%%%%%%%%%%%%%%%%%%%%%%%%%%%
\subsection{Temporal structure of fast flares}
\label{sec:flares:sub:flare_structure}
The flaring epochs display diverse temporal morphologies in both the optical and $\gamma$-ray bands (see Fig.~\ref{fig:LightCurves_flaring_episodes}), ranging from isolated events to extended phases of complex activity, in agreement with previous multi-wavelength campaigns \citep[e.g.,][]{Bonning_3C_opt_gamma_correlation_2009, abdo_3c454_fermi_flare2010_2011}.
Several epochs (E13, E15, E16, E18) feature one or a few major flares superimposed on a plateau emission state.
Epoch E13 displays a fast-rise/exponential-decay (FRED) profile with a nearly constant $\mathrm{CD}$, whereas E16 exhibits the opposite asymmetry (anti-FRED) along with a rapid orphan $\gamma$-ray flare.
Even flares with similar temporal profiles, such as E13 and E18 may present different spectral properties (namely $\eta\simeq 1$ and $\eta \simeq 2$ respectively, and opposite loop sense in index-flux diagrams, see Sect.~\ref{sec:spectra}), and thus likely originating from different mechanisms.
Other epochs exhibit sustained activity with multiple outbursts; E08B is characterised by a prolonged post-flare decay punctuated by short correlated flares, while E09B and E14 remain in extended high states with recurrent activity.

The flare morphology can also differ substantially between energy bands, as reported for previous events in 3C~454.3 \citep[e.g.,][]{gupta_3C454.3flare09_2017MNRAS.472..788G}.
During E09A, the $\gamma$-ray flux increases smoothly, whereas the optical light curve shows multiple peaks, including a rapid flare with a doubling time under $\SI{6}{\hour}$.
Fast flares ($<6$ h) with weak or absent counterparts are detected in both $\gamma$-rays (E10B, E14, E15, E16) and the optical band (E07A, E07B, E09A, E10B, E22).
Notably, heightened and delayed $\gamma$-ray activity seems to be present $7-10$ d after these fast optical flares, a behaviour in principle compatible with mirror models, although detailed modelling is required to test the validity of this scenario in the observed flares.
The observation of fast flares demonstrates that different emission components can decouple within the same flaring epoch, though such behaviour does not necessarily require distinct emission mechanisms \citep[e.g.,][]{rajput_3C454_temporalcorrelation_2019, zhang_simulations_reconnection_2022ApJ...924...90Z}.

\begin{figure}
   \centering
   \resizebox{\hsize}{!}{\includegraphics{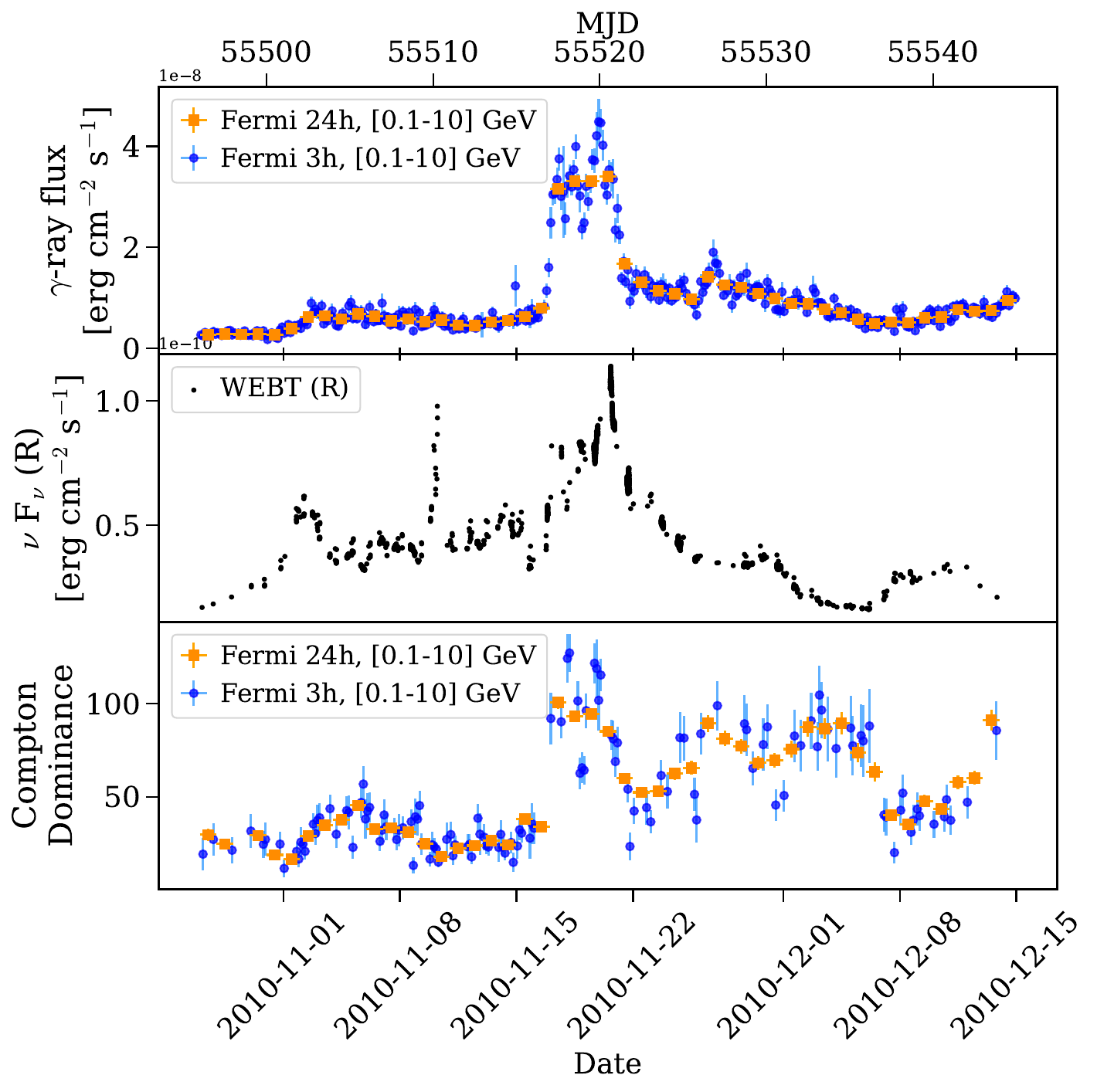}}
    \caption{Light curves and Compton dominance of 3C~454.3 at $\SI{24}{\hour}$ and $\SI{3}{\hour}$ time scales during epoch E10B.}
 \label{fig:LightCurve_2010B}
\end{figure}

\begin{figure}
   %\sidecaption
   %\includegraphics[keepaspectratio, width=12cm]{Images/correlation_24h_3h_2010B.pdf}
   \centering
   \resizebox{\hsize}{!}{\includegraphics{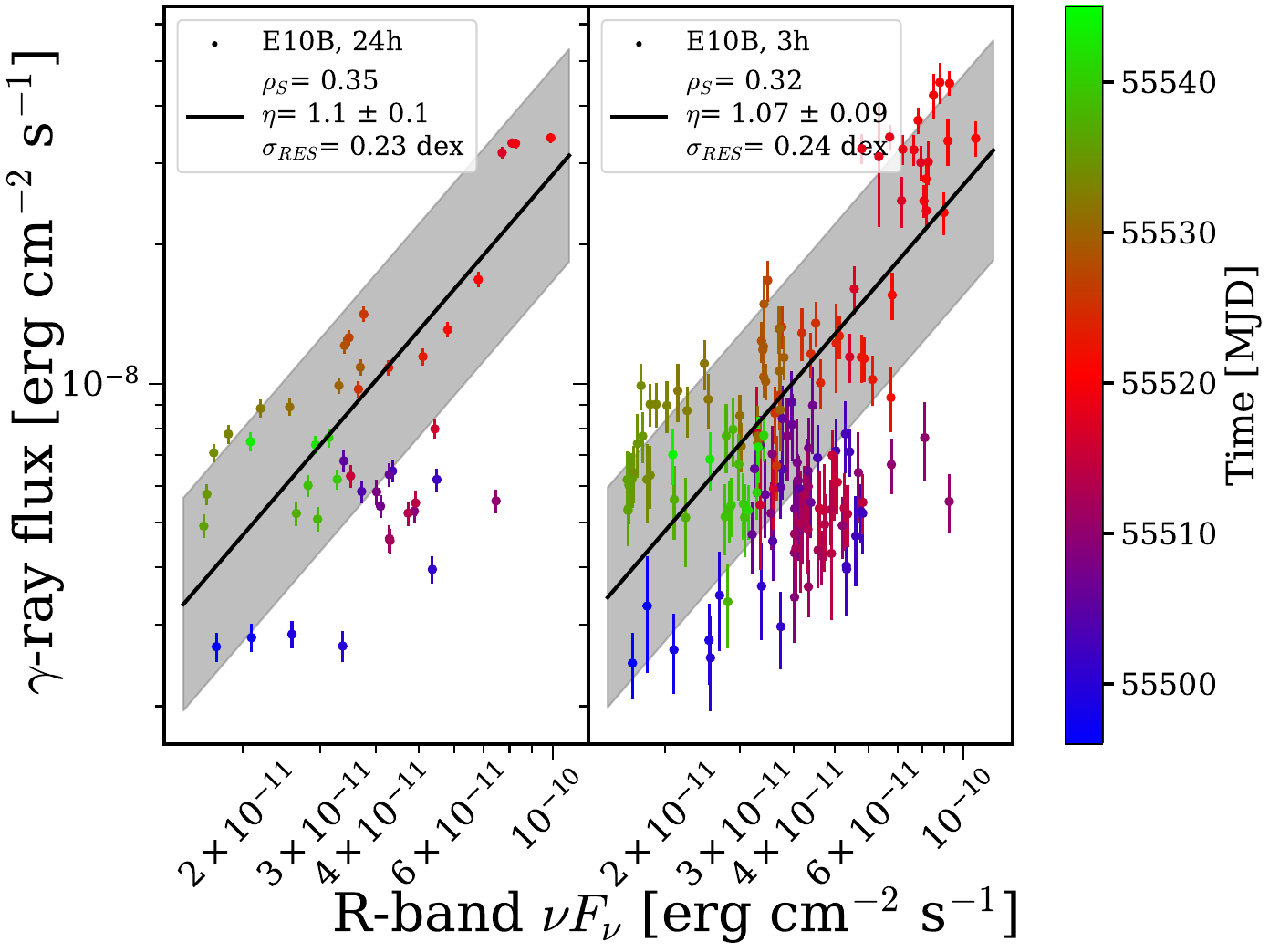}}
   \caption{Correlation between $\gamma$-ray and optical flux of 3C~454.3 for epoch E10B, using $\SI{24}{\hour}$ (left) and $\SI{3}{\hour}$ (right) binning. The shorter binning emphasises the complex structure of the epoch, with points clustered in specific structures rather than scattered uniformly across the average correlation trend.}
 \label{fig:Correlation_2010B:3h}
\end{figure}

Fine temporal resolution is crucial for resolving flare morphologies, as demonstrated by the exceptionally bright epoch E10B, where the $\gamma$-ray light curve can be reconstructed down to $\SI{3}{\hour}$ bins (Fig.~\ref{fig:LightCurve_2010B}).
While the daily light curve suggests a single smooth outburst lasting about four days, the $\SI{3}{\hour}$ resolution reveals a sequence of rapidly evolving sub-flares.
The peak contains multiple intensity maxima accompanied by rapid $\mathrm{CD}$ variations, which drop from $\sim 100$ to $\sim 60$ within $\SI{3}{\hour}$ before recovering equally fast.
Such large $\mathrm{CD}$ variations on intra-day timescales challenge standard one-zone models \citep[e.g.][]{tavani_flaremodel_2015}.
At this same fine resolution, the optical-$\gamma$ correlation (Fig.~\ref{fig:Correlation_2010B:3h}) highlights these substructures, with data points forming separate clusters rather than a single correlation trend.
Orphan optical and orphan $\gamma$-ray flares create nearly horizontal and vertical branches, respectively, while the pre- and post-flare states occupy distinct regions of the parameter space despite having comparable optical fluxes.
These features are largely smoothed out on daily timescales, supporting a scenario where the global outburst is produced by a succession of localised dissipation episodes or multiple emitting regions rather than a single stationary zone \citep[e.g.,][]{nalewajko_flare_loops_2013MNRAS.430.1324N, aharonian_ultrafastvariability_2017ApJ...841...61A}.
The variability of 3C~454.3 cannot thus be captured by a single profile, as even apparently simple outbursts can conceal a rich internal temporal structure at shorter timescales.

%%%%%%%%%%%%%%%%%%%%%%%%%%%%%%%%%%%%%%%%%%
%%%%%%%%%%%%%%%%%%%%%%%%%%%%%%%%%%%%%%%%%%

%%%%%%%%%%%%%%%%%%%%%%%%%%%%%%%%%%%%%%%%%%
%%%%%%%%%%%%%%%%%%%%%%%%%%%%%%%%%%%%%%%%%%

%%%%%%%%%%%%%%%%%%%%%%%%%%%%%%%%%%%%%%%%%%%%%
\subsection{Orphan flares}
\label{sec:flares:sub:orphan_flares}
Several epochs exhibit orphan flares or partial decoupling between bands, demonstrating that optical and $\gamma$-ray emissions do not always evolve coherently.
The observation of orphan flares and band-dependent temporal profiles favours multi-zone scenarios or spatially inhomogeneous particle acceleration.
In leptonic models, individual orphan flares are usually explained by introducing sudden variations of the energy density of the field responsible for the synchrotron or IC radiation.
Namely, variations in the magnetic field may produce optical orphans, whereas variations of seed photon fields can produce $\gamma$-ray orphan.
More subtle forms of band decoupling can also emerge, as in epoch E09A (Fig.~\ref{fig:LightCurves_flaring_episodes:sub:09A}), where the optical and $\gamma$-ray emissions track each other during the initial rise but diverge during the decay phase, leading to a progressive increase in $\mathrm{CD}$.

\subsubsection{Gamma-ray orphan flares}
\label{sec:flares:sub:orphan_flares:sub:gamma}
While it is common to observe flares with a heightened $\gamma$-ray response (CD$\gtrsim 50$), the most explicit examples of fast orphan $\gamma$-ray flares are observed during epochs E14, E15 and E16.
In E14, a fast orphan flare is present between the two main peaks, which are instead observed also in the optical band.
In E16 (Fig.~\ref{fig:LightCurves_flaring_episodes:sub:16}), a sharp orphan flare dominates the high-energy activity during the rise of a broader, slower outburst.
Epoch E15 (Fig.~\ref{fig:LightCurves_flaring_episodes:sub:15}) features multiple orphan events superimposed on a moderately correlated baseline, causing rapid spikes in $\mathrm{CD}$.
Orphan $\gamma$-ray flares are explained within leptonic frameworks by sudden variations of the IC seed photon field that can be realised using several jet geometries and/or particle distributions \citep{vittorini_blobmodel_2014, MacDonald_RingofFire_2015ApJ...804..111M, vittorini_3C279_model_2017, sobacchi_orphanflare_2021MNRAS.503..688S, zhang_simulations_reconnection_2022ApJ...924...90Z, bihan_zech_Orphan3C279_2026JHEAp..5300629L}.
Alternatively, they could also indicate hadronic processes, with $\gamma$-ray emission due to proton synchrotron radiation \citep{bottcher_hadronic_orphan_2005ApJ...621..176B, zhang_2013_hadronicandpolarization_ApJ...774...18Z, cerruti_jetprocesses_2020}.

\subsubsection{Optical orphan flares}
\label{sec:flares:sub:orphan_flares:sub:optical}
Epoch E09A includes a rapid flare at MJD 55087 with fine optical substructure that is entirely absent in $\gamma$-rays.
Optical orphan flares appear less frequently in our sample, with the clearest cases being the fast flares observed in E07A, E07B, E09A, E10B and E22.
While this asymmetry matches previous statistical trends \citep{Liodakis_CorrelationsFlaresOrphans_2019ApJ...880...32L}, it is likely amplified by our selection of epochs based on elevated $\gamma$-ray states.
Optical orphan flares may imply localized changes in the synchrotron-emitting plasma, such as variations in the magnetic field or acceleration efficiency, that leave the external Compton component largely unaffected \citep[this type of flares is also referred to as `sterile', e.g.][]{raiteri_blazarvarreview_2025}.

%%%%%%%%%%%%%%%%%%%%%%%%%%%%%%%%%%%%%%%%%%%%%%%%%%%%%%%%%%%%%%%%%%%%%
%%%%%%%%%%%%%%%%%%%%%%%%%%%%%%%%%%%%%%%%%%%%%%%%%%%%%%%%%%%%%%%%%%%%%
%%%%%%%%%%%%%%%%%%%%%%%%%%%%%%%%%%%%%%%%%%%%%%%%%%%%%%%%%%%%%%%%%%%%%
%%%%%%%%%%%%%%%%%%%%%%%%%%%%%%%%%%%%%%%%%%%%%%%%%%%%%%%%%%%%%%%%%%%%%
%%%%%%%%%%%%%%%%%%%%%%%%%%%%%%%%%%%%%%%%%%%%%%%%%%%%%%%%%%%%%%%%%%%%%
%%%%%%%%%%%%%%%%%%%%%%%%%%%%%%%%%%%%%%%%%%%%%%%%%%%%%%%%%%%%%%%%%%%%%
%%%%%%%%%%%%%%%%%%%%%%%%%%%%%%%%%%%%%%%%%%%%%%%%%%%%%%%%%%%%%%%%%%%%%
%%%%%%%%%%%%%%%%%%%%%%%%%%%%%%%%%%%%%%%%%%%%%%%%%%%%%%%%%%%%%%%%%%%%%
%%%%%%%%%%%%%%%%%%%%%%%%%%%%%%%%%%%%%%%%%%%%%%%%%%%%%%%%%%%%%%%%%%%%%
% SPECTRA
\section{Spectral variability of selected flaring epochs}
\label{sec:spectra}
For selected intervals within each flaring epoch spanning both active and quiescent states, we performed dedicated \textit{Fermi} SED fits in the $0.1-30\,\si{\giga\electronvolt}$ range using the log-parabola model of Eq.~\eqref{eq:LP}.
We examined the spectral index $\alpha$ as a function of flux to search for harder-when-brighter (HWB) trends and spectral hysteresis loops.
Integration times range from one to a few days, and best-fit parameters are listed in Table~\ref{table:spectra}.

%%%%%%%%%%%%%%%%%%%%%%%%%%%%%%%%%%%%%%%%%%
\begin{figure}
   \centering
   \resizebox{\hsize}{!}{\includegraphics{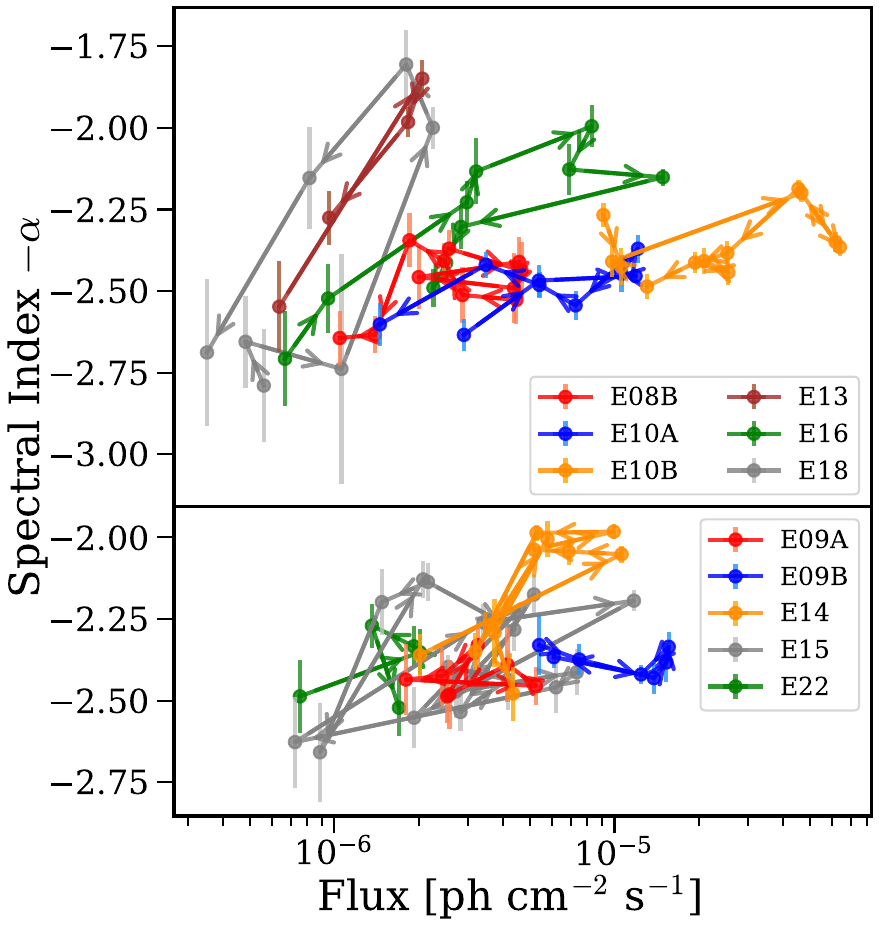}}
   \caption{Evolution of spectral index versus flux for the flaring epochs. Top panel: epochs exhibiting HWB trends and hysteresis loops. Bottom panel: epochs showing no clear correlation.}
 \label{fig:SED_properties:alpha_flux}
\end{figure}
%%%%%%%%%%%%%%%%%%%%%%%%%%%%%%%%%%%%%%%%%%

The index-flux trajectories in Fig.~\ref{fig:SED_properties:alpha_flux} reveal heterogeneous spectral evolution.
Clear or tentative HWB trends occur in some epochs (top panel; E08B, E10A, E10B, E13, E16, E18), while others show no distinct correlation between flux and spectral index (bottom panel; E09A, E09B, E14, E15, E22).
Each epoch occupies a separate region of the parameter space, demonstrating that the brightest flares are not always the hardest.
An HWB trend typically indicates enhanced particle acceleration or fresh electron injection shifting the IC peak to higher energies during bright states. 
By contrast, its absence may reflect competing influences from acceleration, radiative cooling, and variations in the emission region geometry or Doppler factor \citep[e.g.][]{sikora_spectral_trends_2001ApJ...554....1S}.
The directional loops include clockwise (E08B, E13, E16), counter-clockwise (E18), and mixed (E10A, E10B) patterns, with no dominant orientation.
This spectral hysteresis reflects a time lag between spectral and flux evolution from competing acceleration and cooling timescales.
Although the loop morphology depends on the emission process and the electron distribution, clockwise loops generally indicate rapid acceleration followed by slower cooling, whereas counter-clockwise loops can occur when acceleration and cooling timescales are comparable \citep[e.g.][]{kirk_spectral_loops_1998A&A...333..452K,chiaberge_trends_1999MNRAS.306..551C}.

Two representative cases can illustrate HWB trends.
The brightest flare, E10B (Fig.~\ref{fig:spectra:2010B}), hardens significantly from $\alpha=2.60\pm0.06$ to $2.20\pm0.03$ during peak emission, while the curvature $\beta$ remains consistent within errors ($\beta\simeq0.15$).
The CD increases by a factor of four in less than one day to $\sim 100$ (Fig.~\ref{fig:LightCurves_flaring_episodes:sub:10B}).
This behaviour, alongside the structured light curve and hysteresis, suggests that the main flare comprises multiple superimposed acceleration and cooling episodes.
In contrast, the isolated flare E13 (Fig.~\ref{fig:spectra:2013}) features a simple temporal profile (a single flare over a constant plateau, correlated across bands) but exhibits the strongest spectral evolution of the sample.
The index $\alpha$ hardens from $2.55\pm0.10$ to $1.85\pm0.06$, while $\beta$ decreases from $0.27\pm0.09$ to $\lesssim 0.05$.
Despite this spectral variation, the CD remains stable at $\lesssim10$, typical of low states (Fig.~\ref{fig:LightCurves_flaring_episodes:sub:13}).
This indicates that spectral hardness and CD can decouple, probing different aspects of the electron distribution.
We note that this conclusion stems from the observational definition of CD at fixed energies; broadband modelling of peak frequency shifts might modify this interpretation.

%%%%%%%%%%%%%%%%%%%%%%%%%%%%%%%%%%%%%%%%%%
\begin{figure}
   \centering
   \resizebox{\hsize}{!}{\includegraphics{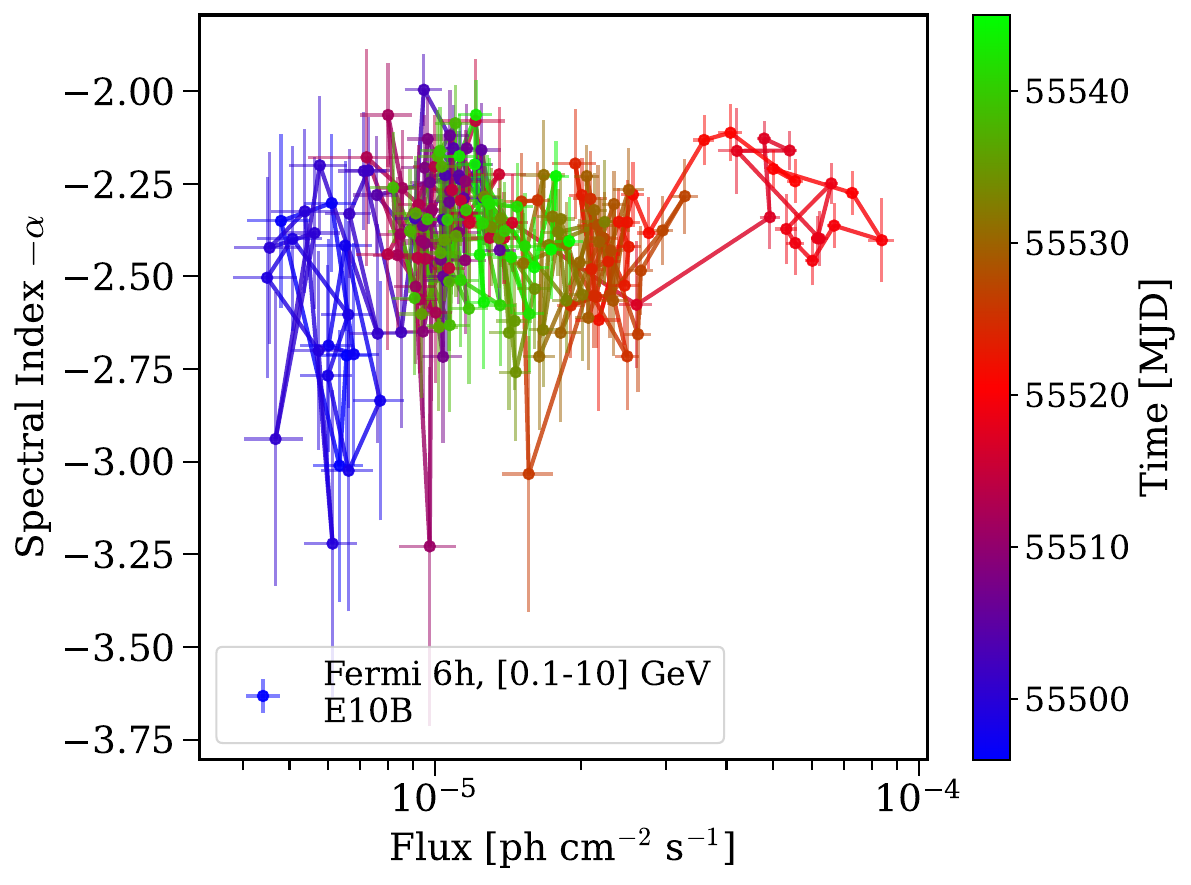}}
    \caption{Evolution of spectral index and flux for the $\SI{6}{\hour}$ binned data of epoch E10B.
    Short integration time reveals complex spectral variability.}
 \label{fig:SED_properties:alpha_flux_2010B}
\end{figure}
%%%%%%%%%%%%%%%%%%%%%%%%%%%%%%%%%%%%%%%%%%

While the day-scale SED sequence for E10B outlines its average evolution, its brightness allows a intra-day analysis. Figure~\ref{fig:SED_properties:alpha_flux_2010B} shows the index-flux track from the $\SI{6}{\hour}$ light curve.
The global trajectory matches the day-scale loop, but the finer sampling reveals repeated loop crossings and reversals, indicating that the smooth trend is a superposition of rapidly changing spectral states.
Although some scatter stems from larger statistical uncertainties, the complex trajectories, particularly during the peak emission at MJD 55517-55521, point to intrinsic intra-day spectral variability, consistent with prior sub-sample studies \citep[e.g.][]{nalewajko_flare_loops_2013MNRAS.430.1324N,meyer_fermilat_variability_2019}.

In summary, 3C~454.3 displays neither a universal HWB trend nor a preferred hysteresis direction.
Comparing E10B and E13 demonstrates that similar spectral hardenings can occur alongside entirely different CD and light-curve profiles, while the intra-day analysis of E10B shows that smooth trends can mask rapid spectral changes.
The relative impact of particle acceleration, radiative cooling, and geometric effects varies substantially between outbursts, reinforcing the heterogeneous variability patterns established in Sect.~\ref{sec:flares}.

%%%%%%%%%%%%%%%%%%%%%%%%%%%%%%%%%%%%%%%%%%%%%%%%%%%%%%%%%%%%%%%%%%%%%
%%%%%%%%%%%%%%%%%%%%%%%%%%%%%%%%%%%%%%%%%%%%%%%%%%%%%%%%%%%%%%%%%%%%%
%%%%%%%%%%%%%%%%%%%%%%%%%%%%%%%%%%%%%%%%%%%%%%%%%%%%%%%%%%%%%%%%%%%%%
%%%%%%%%%%%%%%%%%%%%%%%%%%%%%%%%%%%%%%%%%%%%%%%%%%%%%%%%%%%%%%%%%%%%%
%%%%%%%%%%%%%%%%%%%%%%%%%%%%%%%%%%%%%%%%%%%%%%%%%%%%%%%%%%%%%%%%%%%%%
%%%%%%%%%%%%%%%%%%%%%%%%%%%%%%%%%%%%%%%%%%%%%%%%%%%%%%%%%%%%%%%%%%%%%
%%%%%%%%%%%%%%%%%%%%%%%%%%%%%%%%%%%%%%%%%%%%%%%%%%%%%%%%%%%%%%%%%%%%%
%%%%%%%%%%%%%%%%%%%%%%%%%%%%%%%%%%%%%%%%%%%%%%%%%%%%%%%%%%%%%%%%%%%%%
%%%%%%%%%%%%%%%%%%%%%%%%%%%%%%%%%%%%%%%%%%%%%%%%%%%%%%%%%%%%%%%%%%%%%
\section{Discussion}
\label{sec:discussion}
We discuss in this section the main results of Sects.~\ref{sec:global_lightcurve},~\ref{sec:flares} and~\ref{sec:spectra}.

%%%%%%%%%%%%%%%%%%%%%%%%%%%%%%%%%%%%%%%%%%%%%%%%%%%%%%%%%%%%%%%%%%%%%
\subsection{A hierarchy of complexity}
\label{sec:discussion:sub:complexity}
The analysis of the optical and $\gamma$-ray variability of 3C~454.3 demonstrates that no single emission model characterises the source behaviour across all timescales.
For the full 17-year dataset (Sect.~\ref{sec:global_lightcurve}), the $\gamma$-ray/optical correlation is strong ($\rho_S=0.80-0.87$), in agreement with previous studies \citep{Bonning_3C_opt_gamma_correlation_2009, cohen_blazar_correlations_2014, majumder_mwl_variability_2019, nalewajko_correlation3C454.3_2019A&A...631A...4N}.
However, it features a large dispersion ($\sigma_\text{RES}=0.24$) and a best-fit slope ($\eta\approx1.6-1.7$) incompatible with canonical one-zone predictions for density-driven variability ($\eta=1$ for EC, $\eta=2$ for SSC).
This indicates that the global correlation does not reflect a single radiative mechanism.

Decomposing the sample into individual flaring epochs on timescales $\leq \SI{24}{\hour}$ (Sect.~\ref{sec:flares}) demonstrates that this discrepancy is not merely a statistical blending of two regimes.
Correlation strengths range from $\rho_S\approx0.20$ to $0.81$, showing that the correlation is epoch-dependent rather than guaranteed.
Best-fit slopes range from sub-linear to near-quadratic ($\eta \approx0.4$ to $2.0$), and the CD evolution follows four distinct patterns: stable-with-excursions, asynchronous, gradual, and highly structured.
Flare morphologies vary similarly, including FRED and anti-FRED profiles, plateaus with single or multiple superimposed flares, and band-decoupled substructures.
Orphan flares occur in both bands across multiple epochs, matching studies that find orphan events to be more common than previously thought \citep{Liodakis_CorrelationsFlaresOrphans_2019ApJ...880...32L, dejaeger_blazarflare_correlation_2023}.
Each epoch appears governed by distinct physical drivers rather than variations around a common mean.

Spectral analysis (Sect.~\ref{sec:spectra}) confirms that this diversity persists at the highest temporal resolution.
HWB trends and hysteresis loops occur in some epochs but are absent in others, with no preferred loop direction.
Comparing E10B and E13 shows that two epochs with substantial spectral hardening can exhibit entirely different CD evolution: E10B reaches $\text{CD} \sim 100$ while E13 remains at quiescent levels ($\lesssim 10$).
This demonstrates that spectral index and Compton dominance can track different properties of the emitting particle population and seed photon field properties.
For E10B, the spectral evolution appearing as a single loop at day-scale resolution resolves at $\SI{6}{\hour}$ scales into crossing and reversing trajectories (Figs.~\ref{fig:SED_properties:alpha_flux} and~\ref{fig:SED_properties:alpha_flux_2010B}).
This substructure cannot be explained by statistical uncertainties alone, consistent with prior sub-sample results \citep{nalewajko_flare_loops_2013MNRAS.430.1324N, meyer_fermilat_variability_2019}.

Thus, the failure of standard one-zone models is not restricted to orphan flares or extreme states; it affects the global baseline, becomes epoch-specific when events are isolated, and intensifies with finer time resolutions.
The underlying complexity compounds rather than averaging out at shorter timescales.

%%%%%%%%%%%%%%%%%%%%%%%%%%%%%%%%%%%%%%%%%%%%%%%%%%%%%%%%%%%%%%%%%%%%%
\subsection{Physical interpretation of the Compton dominance}
\label{sec:discussion:sub:CD}
The bolometric CD provides physical insights that can help understand the observations.
We consider a spherical plasmoid of radius $l'$, particle density $n'$, and rest-frame particle Lorentz factor $\gamma'$, with primed quantities denoting measurements in its rest frame.
The SSC contribution to the CD scales as
\begin{equation}
    \label{eq:CD_SSC:estimate}
    CD_\text{SSC}
    \simeq 0.5 \, \frac{n'}{\SI{1E2}{\centi\meter^{-3}}}\,\frac{l'}{\SI{1E16}{\centi\meter}}\,\left(\frac{\gamma'}{10^3}\right)^2,
\end{equation}
while EC scattering off disc luminosity $L_D$ in an isotropic BLR gives
\begin{equation}
    \label{eq:CD_EC_BLR:estimate}
    CD_\text{EC}
    \simeq 27 \cdot \frac{a}{0.2} \frac{f}{0.15} \frac{L_D}{\SI{1E45}{\erg\per\second}} \left(\frac{\Gamma}{10}\right)^2 \left(\frac{B'}{\SI{1}{\gauss}}\right)^{-2} \left(\frac{R_\text{BLR}}{\SI{1E17}{\centi\meter}}\right)^{-2},
\end{equation}
where $a$ is the cloud-system coverage and $f$ the single-cloud reflectivity, meaning a fraction $a\cdot f$ of the disc luminosity $L_D$ is reflected by the BLR clouds \citep{dermer_basicequations_1997ApJS..109..103D, sikora_EC_1994ApJ...421..153S, vittorini_blobmodel_2014}.

For typical FSRQ parameters, SSC alone yields $\mathrm{CD}\lesssim1$, whereas steady EC from an isotropic BLR gives $\mathrm{CD}\sim20-30$.
This prediction roughly matches the off-flare baseline CD distribution in Fig.~\ref{fig:ComptonDominance} ($\mathrm{CD}\approx5-10$), accounting for dilution from the thermal disc contribution to the $R$ band \citep{raiteri+_2007A&A...473..819R, raiteri_3C454_optical_2011, Amaya_States3C454.3_2021ApJ}.
However, this steady EC-BLR framework cannot reproduce the high-CD tail ($\mathrm{CD}\gtrsim40-120$) observed during flares, nor the rapid timescales often observed (Fig.~\ref{fig:LightCurves_flaring_episodes}).
The natural variability timescale of the BLR photon field is set by $L_D$, which varies over months to years; even with relativistic compression, this yields $\Delta t\sim\mathrm{days}$ at best, failing to explain the $\lesssim \SI{3}{\hour}$ CD excursions in E10B (Fig.~\ref{fig:LightCurve_2010B}) and $\lesssim \SI{6}{\hour}$ in other flaring epochs.

At least two distinct mechanisms can account for these rapid, high-CD excursions \citep{vittorini_blobmodel_2014}.
First, a plasmoid moving through a discrete BLR cloud distribution can closely approach (`fly-by') a single cloud, locally boosting the seed photon density by a factor 
$
\displaystyle f_\text{fly-by} = 1+ a^{-1} \left( \frac{R_\text{cl}}{2\,b}\right)^2 \simeq 2-3
$, where $R_\text{cl}$ is the cloud radius and $b$ the impact parameter ($R_\text{cl}\sim b \sim 10^{16}$ cm).
A fly-by can raise the CD to $\sim40-60$ on a timescale $
\displaystyle \Delta t \simeq \SI{2}{\hour} \left( \frac{b}{\SI{1E16}{\centi\meter}} \right) \, \left( \frac{\Gamma}{10} \right)^{-2}
$, matching most CD peaks in Fig.~\ref{fig:LightCurves_flaring_episodes} but not the most extreme values.
This scenario also predicts an asymmetric flare profile (gradual rise as $\displaystyle \Gamma^2/b^2(t)$, abrupt decay as $\displaystyle \Gamma^{-2}$), reproducing anti-FRED shapes like that in E13 (Fig.~\ref{fig:LightCurves_flaring_episodes:sub:13}).
Second, mirror models (where the synchrotron field is re-scattered by a BLR cloud or a second jet component)can boost the CD to $\sim100$.
These introduce a day-scale delay between the synchrotron and IC responses, offering an explanation for delayed correlations and asynchronous evolution \citep{vittorini_3C279_model_2017, bottcher_MirrorModels_2021Physi...3.1112B}.

%%%%%%%%%%%%%%%%%%%%%%%%%%%%%%%%%%%%%%%%%%%%%%%%%%%%%%%%%%%%%%%%%%%%%
\subsection{Correlation slope and variability drivers}
\label{sec:discussion:sub:slope}
The $\gamma$-ray/optical flux correlation slope $\eta$ depends on both the radiative mechanism (SSC or EC) and the physical parameter driving the variability.
In a standard one-zone model, the synchrotron and IC luminosities scale as:
\begin{align}
    & L_\text{syn}    \propto {n'}   \, {l'}^3 \, {\gamma'}^2 \, {B'}^2 \, \delta^4 \label{eq:proportionality:syn} \\
    & L_\text{SSC}    \propto {n'}^2 \, {l'}^4 \, {\gamma'}^4 \, {B'}^2 \, \delta^4 \label{eq:proportionality:SSC} \\
    & L_\text{EC-BLR} \propto {n'}   \, {l'}^3 \, {\gamma'}^2 \, U_\text{BLR} \, \delta^6,\label{eq:proportionality:ECBLR}
\end{align}
where $B'$ is the rest-frame magnetic field, $U_\text{BLR}$ the external field energy density and $\delta$ is the bulk Doppler factor \citep{dermer_basicequations_1997ApJS..109..103D, bottcher_BLLACs_2002ApJ...581..127B}.
These scaling relations predict: orphan optical flares ($\eta=0$) for EC variability driven by $B'$; linear correlations ($\eta=1$) for SSC driven by $B'$ or EC driven by $n'$, $\gamma'$, or $l'$; super-linear correlations for SSC driven by $l'$ ($\eta=4/3$) or EC driven by $\delta$ ($\eta=3/2$); and quadratic scaling ($\eta=2$) for SSC driven by $n'$ or $\gamma'$.

While this framework seems to offer simple order-of-magnitude explanations for the different epoch-resolved slopes in Sect.~\ref{sec:flares:sub:correlation_slope}, the temporal and spectral complexity and the CD values observed, often disfavours applying one-zone approximations across an entire epoch, exceeding a simple binary EC/SSC classification.
Crucially, the global slope $\eta\approx1.65$ \citep[consistent with][]{bonnoli_3C454.3Correlation09_2011MNRAS.410..368B, chatterjee_Similarity_2012ApJ...749..191C} overlaps with multiple scenarios simultaneously but matches none individually.
This supports the conclusion that separate epochs are governed by distinct combinations of mechanisms and physical drivers rather than a single global process.

%%%%%%%%%%%%%%%%%%%%%%%%%%%%%%%%%%%%%%%%%%%%%%%%%%%%%%%%%%%%%%%%%%%%%
\subsection{Constraints from fast variability}
\label{sec:discussion:sub:fast}
Our systematic analysis on timescales $\leq\SI{24}{\hour}$ extends the $\gamma$-ray/optical correlation into the fast-variability regime.
While the strong global correlation \citep{cohen_blazar_correlations_2014, majumder_mwl_variability_2019, Kushwaha_3C454.3Flare14_2017MNRAS.464.2046K, nalewajko_correlation3C454.3_2019A&A...631A...4N} is confirmed, its limitations emerge at short timescales due to significant scatter.
The epoch-dependent correlation strengths, slopes, CD values, and spectral tracks can explain why diverse emission frameworks—including one-zone and multi-zone EC, SSC, and mixed models—have been proposed for this source \citep{Vercellone_3C454flare2010_2011, abdo_3c454_fermi_flare2010_2011, wehrle_3C454.3Flare10Conic_2012ApJ...758...72W, jorstad_3C454.3Flares0910_2013ApJ...773..147J, coogan_3C454.3Flare14_2016MNRAS.458..354C, weaver_3C454_june2016_2019, Palafox_3C454.3RadioGamma_2025A&A...696A..70P}.

The combination of orphan flares, variable slopes, and CD evolution tracks points to a jet where localised dissipation, magnetic reconnection, plasmoid-cloud interactions, or moving sub-regions intermittently dominate the emission \citep{tavani_flaremodel_2015, vittorini_3C279_model_2017, aharonian_ultrafastvariability_2017ApJ...841...61A, sobacchi_orphanflare_2021MNRAS.503..688S, zhang_simulations_reconnection_2022ApJ...924...90Z}.
The persistence of distinct, sometimes opposing behaviours across epochs separated by years indicates that this structured environment is an intrinsic property of the jet rather than a transient feature of individual outbursts.

Fast variability can also constrain the emission region geometry and the central black hole mass.
The rapid $t_\text{var} \lesssim \SI{3}{\hour}$ flux doubling during the E10B super-flare (Fig.~\ref{fig:LightCurve_2010B}) provides a causality-based upper limit on the emission region size.
Comparing this timescale to the Schwarzschild time yields $
\displaystyle 
    M_\bullet \lesssim \frac{(1+z) \, c^3 \, t_\text{var}}{2\, G \, \delta}
    \simeq 2 \times10^8\,M_\odot \left(\frac{t_\text{var}}{3 \text{ h}}\right) \left( \frac{\delta}{10}\right)^{-1}
$ \citep{dermer_constraints_2008, levinson_constraints_2008}.
This estimate falls below the measured mass $M_\bullet=10^{8.83}\,M_\odot$ \citep{Gu_2001_BHMASS_MNRAS.327.1111G}.
This tension in fast-variable blazars is typically interpreted as a breakdown of simple one-zone causality arguments \citep{salvati_mkn421_variability_1998, Rieger_microvariabilityreview_2019}, requiring compact, localised jet substructures as the primary emission sites \citep{giannios_jets_in_jet_2009, levinson_flaresFromSMBHgaps_2011ApJ...730..123L}.

%%%%%%%%%%%%%%%%%%%%%%%%%%%%%%%%%%%%%%%%%%%%%%%%%%%%%%%%%%%%%%%%%%%%%
%%%%%%%%%%%%%%%%%%%%%%%%%%%%%%%%%%%%%%%%%%%%%%%%%%%%%%%%%%%%%%%%%%%%%
%%%%%%%%%%%%%%%%%%%%%%%%%%%%%%%%%%%%%%%%%%%%%%%%%%%%%%%%%%%%%%%%%%%%%
%%%%%%%%%%%%%%%%%%%%%%%%%%%%%%%%%%%%%%%%%%%%%%%%%%%%%%%%%%%%%%%%%%%%%
%%%%%%%%%%%%%%%%%%%%%%%%%%%%%%%%%%%%%%%%%%%%%%%%%%%%%%%%%%%%%%%%%%%%%
%%%%%%%%%%%%%%%%%%%%%%%%%%%%%%%%%%%%%%%%%%%%%%%%%%%%%%%%%%%%%%%%%%%%%
%%%%%%%%%%%%%%%%%%%%%%%%%%%%%%%%%%%%%%%%%%%%%%%%%%%%%%%%%%%%%%%%%%%%%
%%%%%%%%%%%%%%%%%%%%%%%%%%%%%%%%%%%%%%%%%%%%%%%%%%%%%%%%%%%%%%%%%%%%%
%%%%%%%%%%%%%%%%%%%%%%%%%%%%%%%%%%%%%%%%%%%%%%%%%%%%%%%%%%%%%%%%%%%%%
\section{Conclusions}
\label{sec:conclusions}
We presented a systematic analysis of the optical and $\gamma$-ray variability of 3C~454.3 spanning over 17 years, combining AGILE, \textit{Fermi}-LAT, and WEBT $R$-band data at temporal resolutions from $\leq \SI{24}{\hour}$ down to $\SI{3}{\hour}$.
This homogeneous treatment connects long-term statistical behaviour with individual flares.
No single timescale or analysis level converges on a unique emission model; instead, shorter timescales reveal progressively greater complexity.

For the full sample, the $\gamma$-ray/optical correlation is strong ($\rho_S \simeq 0.80-0.87$), matching previous long-term studies \citep{Bonning_3C_opt_gamma_correlation_2009, cohen_blazar_correlations_2014, majumder_mwl_variability_2019, nalewajko_correlation3C454.3_2019A&A...631A...4N}.
However, its slope ($\eta \approx 1.6-1.7$) and large dispersion ($\sigma_\mathrm{RES} \simeq 0.24$) are incompatible with standard one-zone expectations for density-driven variability in EC or SSC scenarios.
The long-term correlation thus represents a statistical superposition of multiple physical regimes rather than a single radiative process.

Decomposing the sample into individual flaring epochs on intra-day timescales resolves this global trend into genuine physical diversity.
Correlation strengths span $\rho_S \approx 0.2-0.8$, slopes range from sub-linear to near-quadratic ($\eta \approx 0.4-2.0$), and the CD exhibits diverse evolution tracks.
These feature FRED and anti-FRED profiles, multi-peaked structures, plateau states, band-decoupled substructures, and orphan flares in both bands \citep{Liodakis_CorrelationsFlaresOrphans_2019ApJ...880...32L, dejaeger_blazarflare_correlation_2023}.
Each epoch is therefore governed by distinct emission conditions rather than a single variability mode.

At the spectral level, HWB trends and hysteresis loops occur intermittently with no preferred loop orientation.
Epochs with similar HWB trends (e.g. E10B and E13) exhibit markedly different CD behaviours, demonstrating that spectral index and bolometric energetics trace partially decoupled properties of the electron population.
intra-day resolution shows that smooth day-scale spectral tracks resolve into rapidly changing, crossing, and reversing trajectories, consistent with prior small-sample studies for this source \citep{nalewajko_flare_loops_2013MNRAS.430.1324N, meyer_fermilat_variability_2019}.

The observation of fast variability features provides independent physical constraints.
The shortest resolved variability timescale ($t_\mathrm{var} \lesssim \SI{3}{\hour}$) implies a highly compact emission region.
This reinforces the failure of standard one-zone causality arguments and points toward models involving localised jet substructures \citep{salvati_mkn421_variability_1998, Rieger_microvariabilityreview_2019, giannios_jets_in_jet_2009, levinson_flaresFromSMBHgaps_2011ApJ...730..123L}.

Overall, our dataset indicates that the strong global optical-$\gamma$ correlation typically reported for 3C~454.3 \citep{cohen_blazar_correlations_2014, majumder_mwl_variability_2019, nalewajko_correlation3C454.3_2019A&A...631A...4N} masks a complex hierarchy of epoch-dependent variability regimes.
This hierarchy points toward a jet environment where localised, transient dissipation processes (such as magnetic reconnection, plasmoid-cloud interactions, or moving substructures) intermittently dominate the emission \citep{tavani_flaremodel_2015, vittorini_3C279_model_2017, aharonian_ultrafastvariability_2017ApJ...841...61A, sobacchi_orphanflare_2021MNRAS.503..688S, zhang_simulations_reconnection_2022ApJ...924...90Z}.
The variability is thus governed by a time-dependent combination of particle injection, radiative cooling, and structured jet dynamics rather than a single homogeneous zone.
Finer temporal resolution reveals progressively richer structure, linking the global correlation and dispersion to the diverse phenomenology of individual flares.

This study highlights the importance of systematic intra-day and multi-epoch monitoring for bright FSRQs, which might be performed using new and upcoming multi-wavelength facilities such as the Cherenkov Telescope Array Observatory \citep[CTAO,][]{CTAO_BOOK_2019scta.book.....C}, the Compton Spectrometer and Imager mission \citep[COSI,][]{tomsick_COSI_2023} and the Legacy Survey of Space and Time (LSST) started on June 29, 2026 at the Rubin Observatory \citep{LSST_Paper_2019ApJ...873..111I}, which will further probe fast blazar variability and its connection to jet microphysics.

%%%%%%%%%%%%%%%%%%%%%%%%%%%%%%%%%%%%%%%%%%%%%%%%%%%%%%%%%%%%%%
% ACKNOWLEDGEMENTS
\begin{acknowledgements}
      The AGILE and Fermi $\gamma$-ray light curves and spectra presented in this work are publicly available to support reproducibility \citep{panebianco_2026_dataset3C454.3}.
      This work is partly based on data taken and assembled by the WEBT collaboration and stored in the WEBT archive at the Osservatorio Astrofisico di Torino - INAF (\url{https://www.oato.inaf.it/blazars/webt/}).
      We acknowledge the SMARTS Consortium for providing the optical and near-infrared observations included in the optical dataset.
      We acknowledge the Fermi-LAT Collaboration for making the publicly available gamma-ray data and software used in this work accessible to the scientific community.
      AGILE is a mission of the Italian Space Agency (ASI), with scientific and programmatic participation by the Italian National Institute for Astrophysics (INAF) and the Italian National Institute for Nuclear Physics (INFN).
      We thank the institutions involved in the past and present production and publication of AGILE data and software
      for making them accessible to the scientific community.
      G. Panebianco acknowledges financial support from the Italian National Institute for Astrophysics (INAF) through the research fellowship 183/2024.
      P. Coppi acknowledges support from NASA Fermi Guest Observer grant 80NSSC22K1614.
      C. M. Raiteri ad M. Villata acknowledge financial support from the INAF Fundamental Research Funding Call 2023 (PI: Raiteri).
\end{acknowledgements}

%%%%%%%%%%%%%%%%%%%%%%%%%%%%%%%%%%%%%%%%%%%%%%%%%%%%%%%%%%%%%%
% BIBLIOGRAPHY
\bibliographystyle{aa} % style aa.bst
\bibliography{references}

%%%%%%%%%%%%%%%%%%%%%%%%%%%%%%%%%%%%%%%%%%%%%%%%%%%%%%%%%%%%%%
% APPENDIX
\begin{appendix}
\onecolumn

\section{Light curves and spectra of selected flaring epochs}
\label{sec:app:flares}
Fig.~\ref{fig:LightCurves_flaring_episodes} shows the light curves of the selected flaring epochs in Table~\ref{table:Periods}.
Additionally, we estimated the spectra of 3C~454.3 during periods covering both flaring and off-flare states (see Table~\ref{table:spectra}).

%%%%%%%%%%%%%%%%%%%%%%%%%%%%%%%%%%%%%%%%%%%%%%%%%%%%%%%%%%%%%%%%%%%%%%%%%%%%%%%%%%%
%%%%%%%%%%%%%%%%%%%%%%%%%%%%%%%%%%%%%%%%%%%%%%%%%%%%%%%%%%%%%%%%%%%%%%%%%%%%%%%%%%%
%%%%%%%%%%%%%%%%%%%%%%%%%%%%%%%%%%%%%%%%%%%%%%%%%%%%%%%%%%%%%%%%%%%%%%%%%%%%%%%%%%%
\begin{figure*}[hb!]
    \centering

    \begin{subfigure}{0.49\textwidth}
        \centering
        \includegraphics[width=\linewidth]{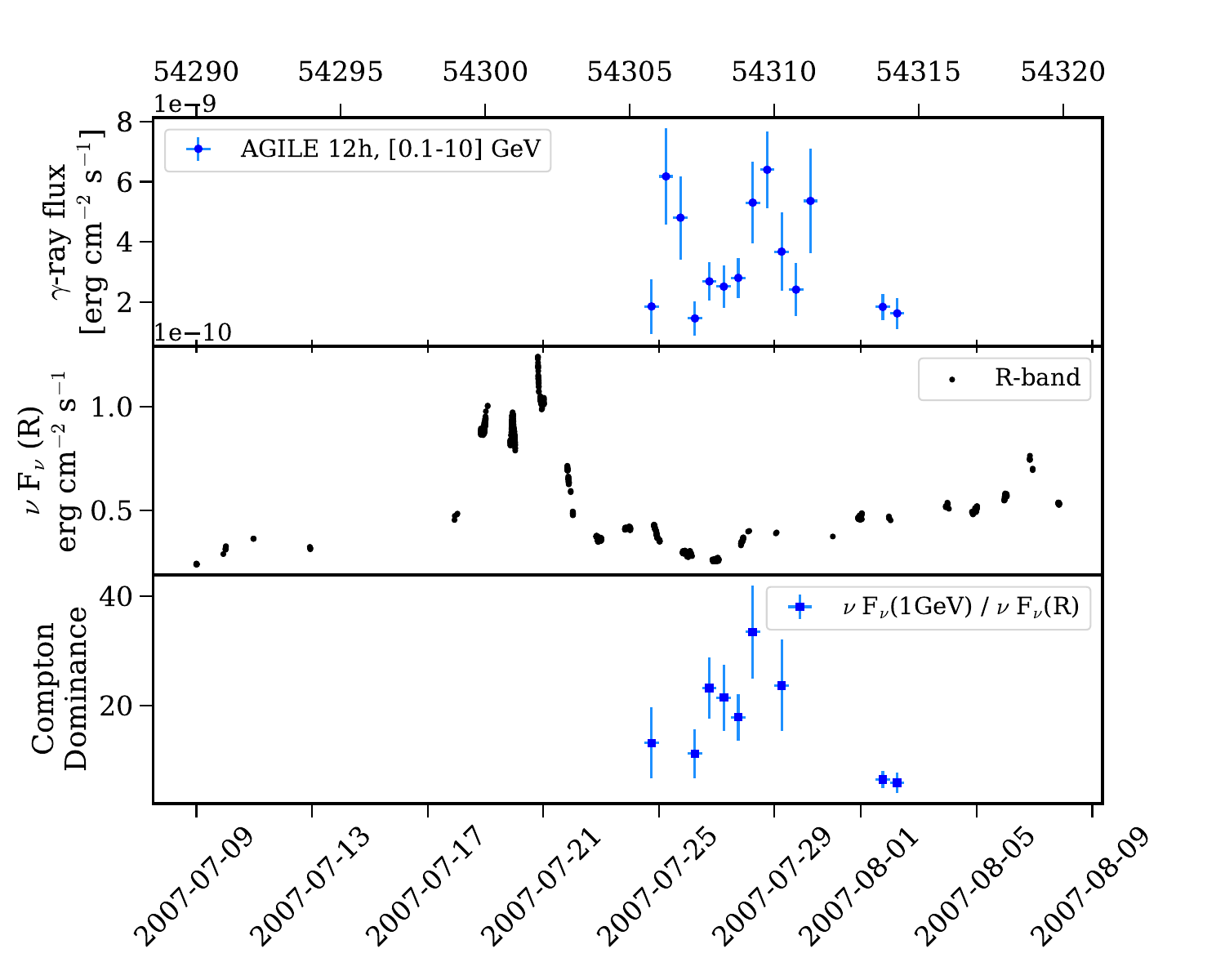}
        \caption{E07A, $12$ h}
        \label{fig:LightCurves_flaring_episodes:sub:07A}
    \end{subfigure}
    \begin{subfigure}{0.49\textwidth}
        \centering
        \includegraphics[width=\linewidth]{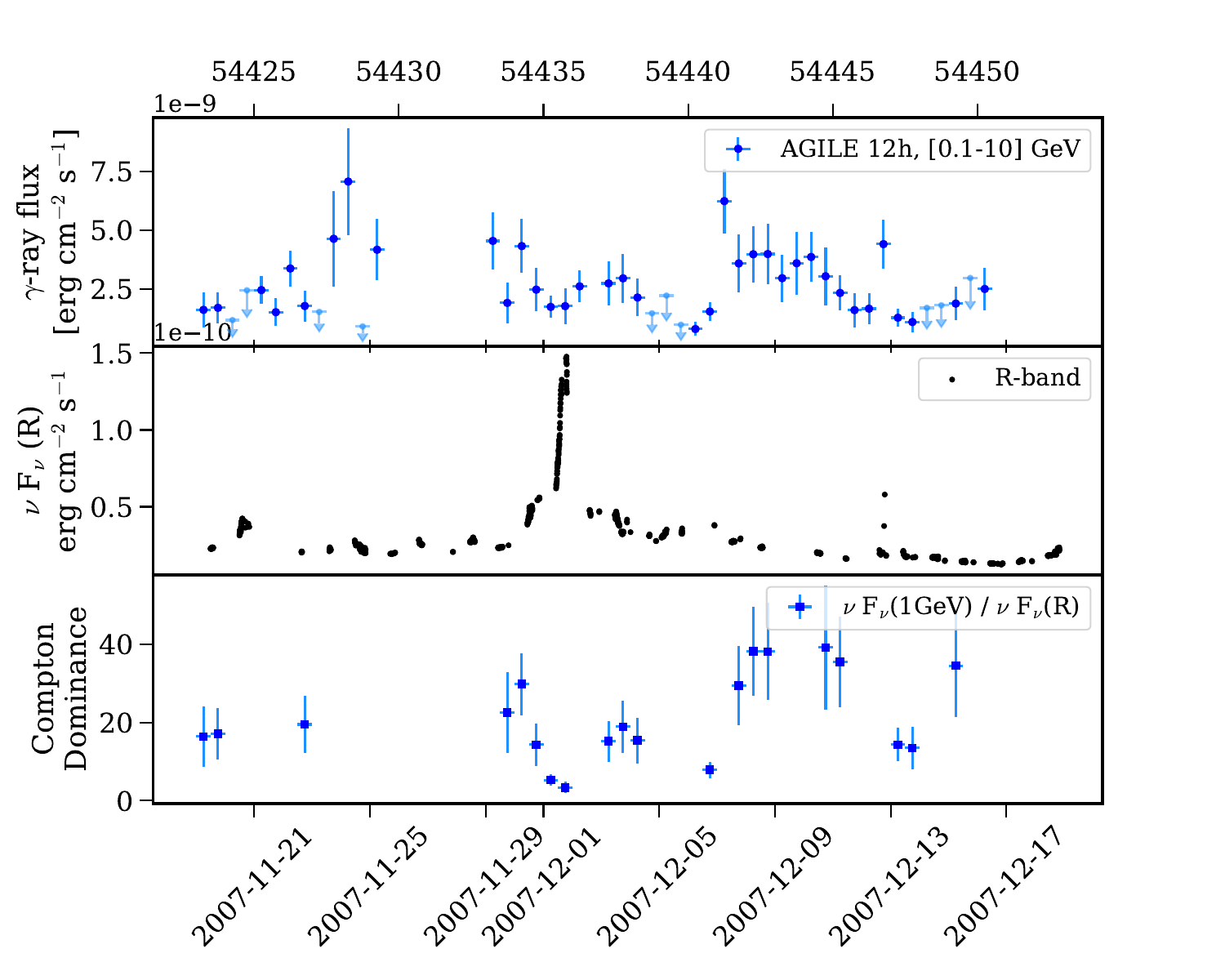}
        \caption{E07B, $12$ h}
        \label{fig:LightCurves_flaring_episodes:sub:07B}
    \end{subfigure}

    \begin{subfigure}{0.49\textwidth}
        \centering
        \includegraphics[width=\linewidth]{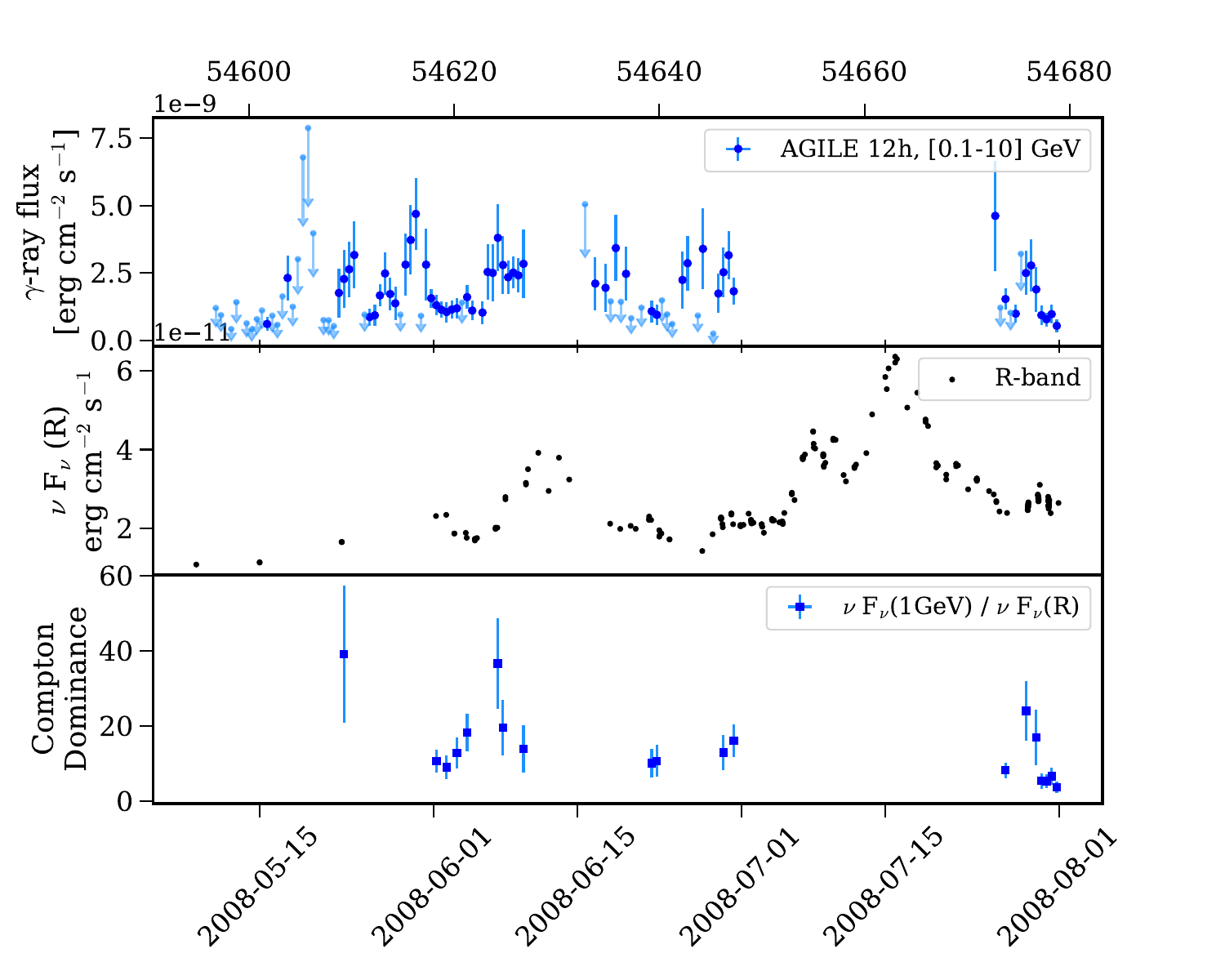}
        \caption{E08A, $12$ h}
        \label{fig:LightCurves_flaring_episodes:sub:08A}
    \end{subfigure}
    \begin{subfigure}{0.49\textwidth}
        \centering
        \includegraphics[width=\linewidth]{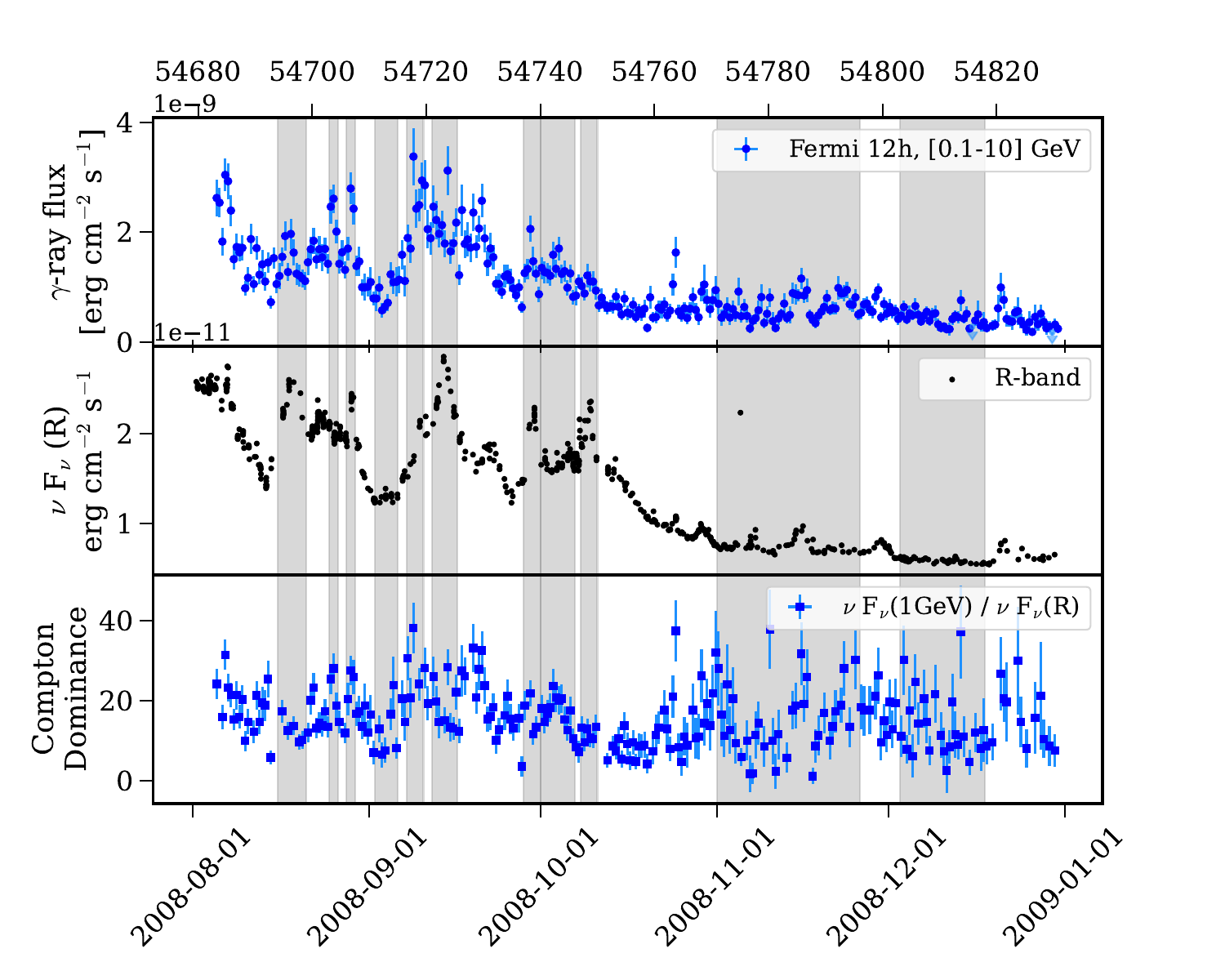}
        \caption{E08B, $12$ h}
        \label{fig:LightCurves_flaring_episodes:sub:08B}
    \end{subfigure}
    
    \caption{Multi-wavelength light curves and Compton dominance evolution during the selected flaring epochs of 3C~454.3 ($\SI{6}{\hour}$ or $\SI{12}{\hour}$ binning). Time is displayed in calendar date (bottom axis) and MJD (top axis). Top panels: $0.1-10~\si{\GeV}$ $\gamma$-ray energy flux from AGILE-GRID (prior to August 2008) and \textit{Fermi}-LAT. Flux points are in blue with light blue error bars. Light blue arrows represent $95\%$ confidence level upper limits. Middle panels: optical $\nu F_\nu$ flux in the $R$ band. Bottom panels: Compton dominance, defined as the ratio between the $\nu F_\nu$ spectral energy distribution values at $\SI{1}{\GeV}$ and at $\nu_\text{eff,R}$. Grey vertical shaded bands indicate the integration intervals used for the spectral analysis in Table~\ref{table:spectra}.}
    \label{fig:LightCurves_flaring_episodes}
\end{figure*}

\begin{figure*}
    \centering
    \ContinuedFloat

    \begin{subfigure}{0.49\textwidth}
        \centering
        \includegraphics[width=\linewidth]{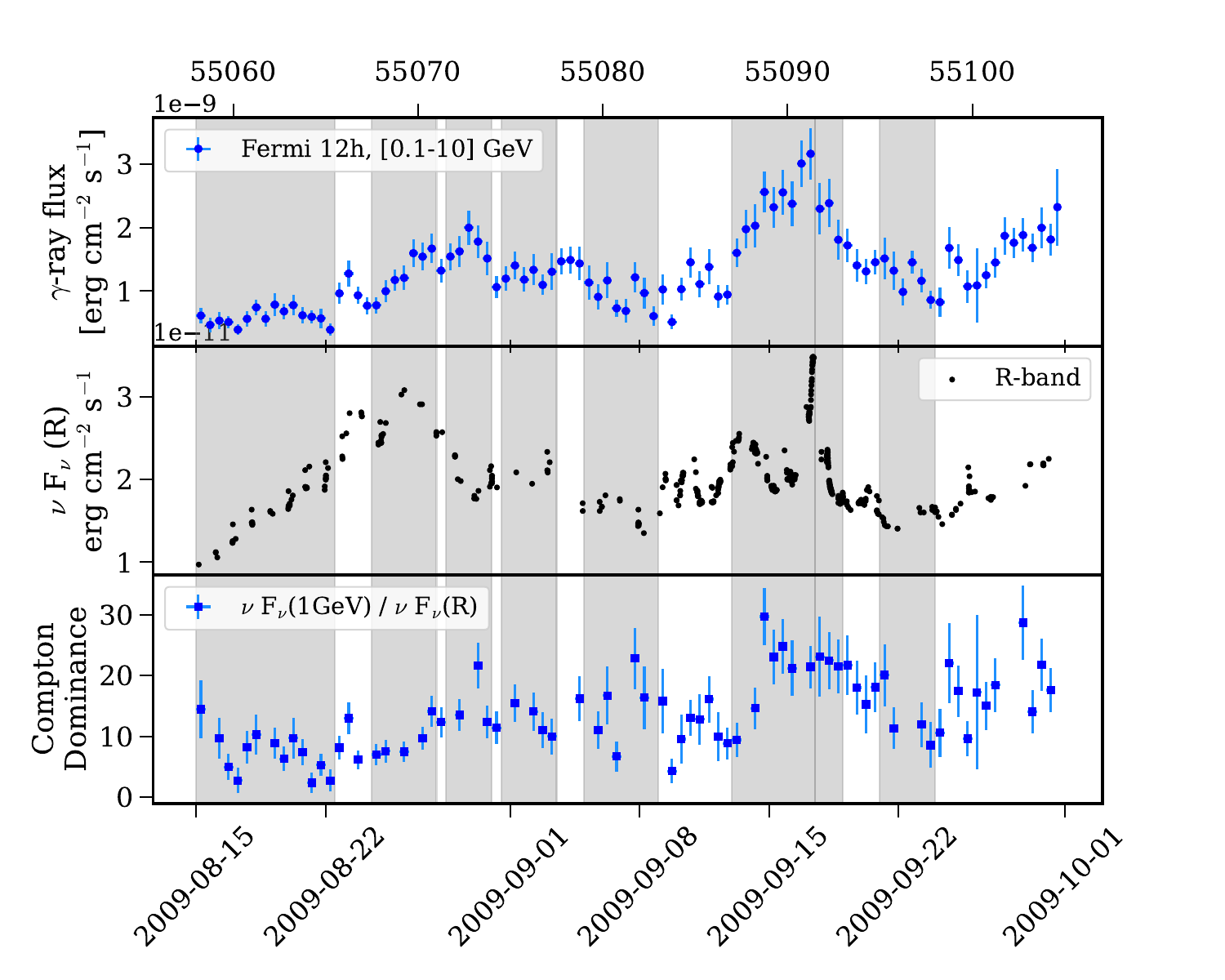}
        \caption{E09A, $12$ h}
        \label{fig:LightCurves_flaring_episodes:sub:09A}
    \end{subfigure}
    \begin{subfigure}{0.49\textwidth}
        \centering
        \includegraphics[width=\linewidth]{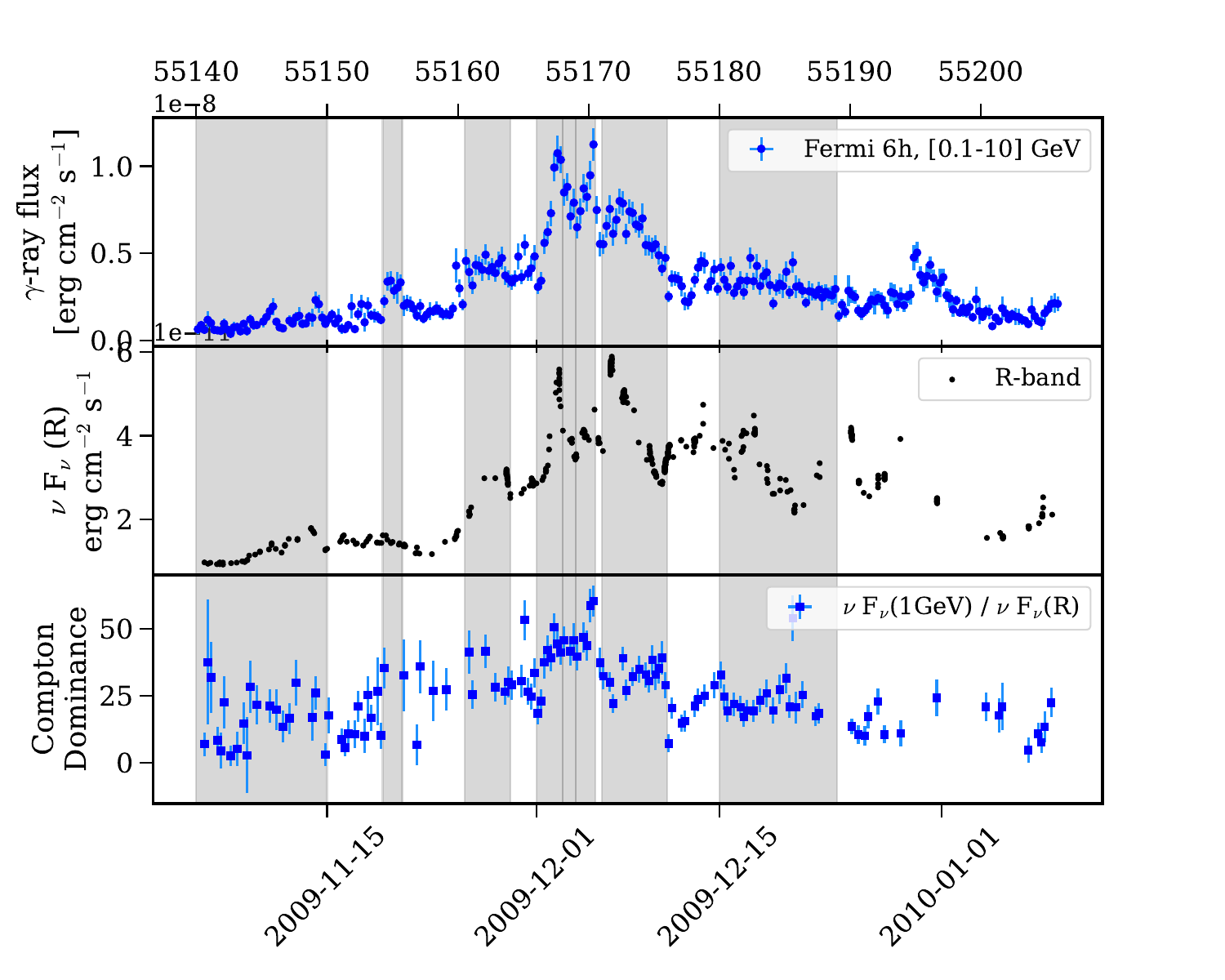}
        \caption{E09B, $6$ h}
        \label{fig:LightCurves_flaring_episodes:sub:09B}
    \end{subfigure}

    \begin{subfigure}{0.49\textwidth}
        \centering
        \includegraphics[width=\linewidth]{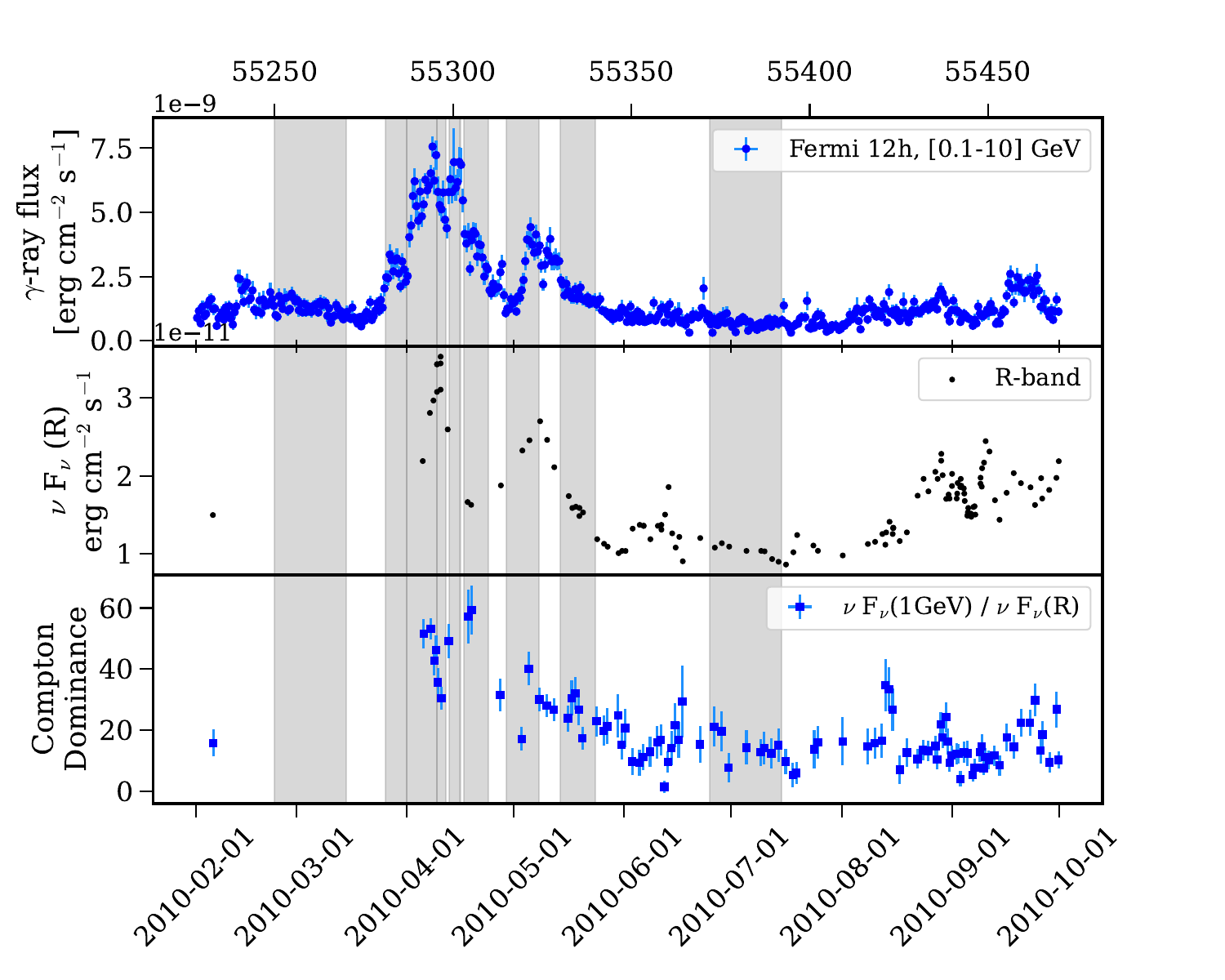}
        \caption{E10A, 12 h}
        \label{fig:LightCurves_flaring_episodes:sub:10A}
    \end{subfigure}
    \begin{subfigure}{0.49\textwidth}
        \centering
        \includegraphics[width=\linewidth]{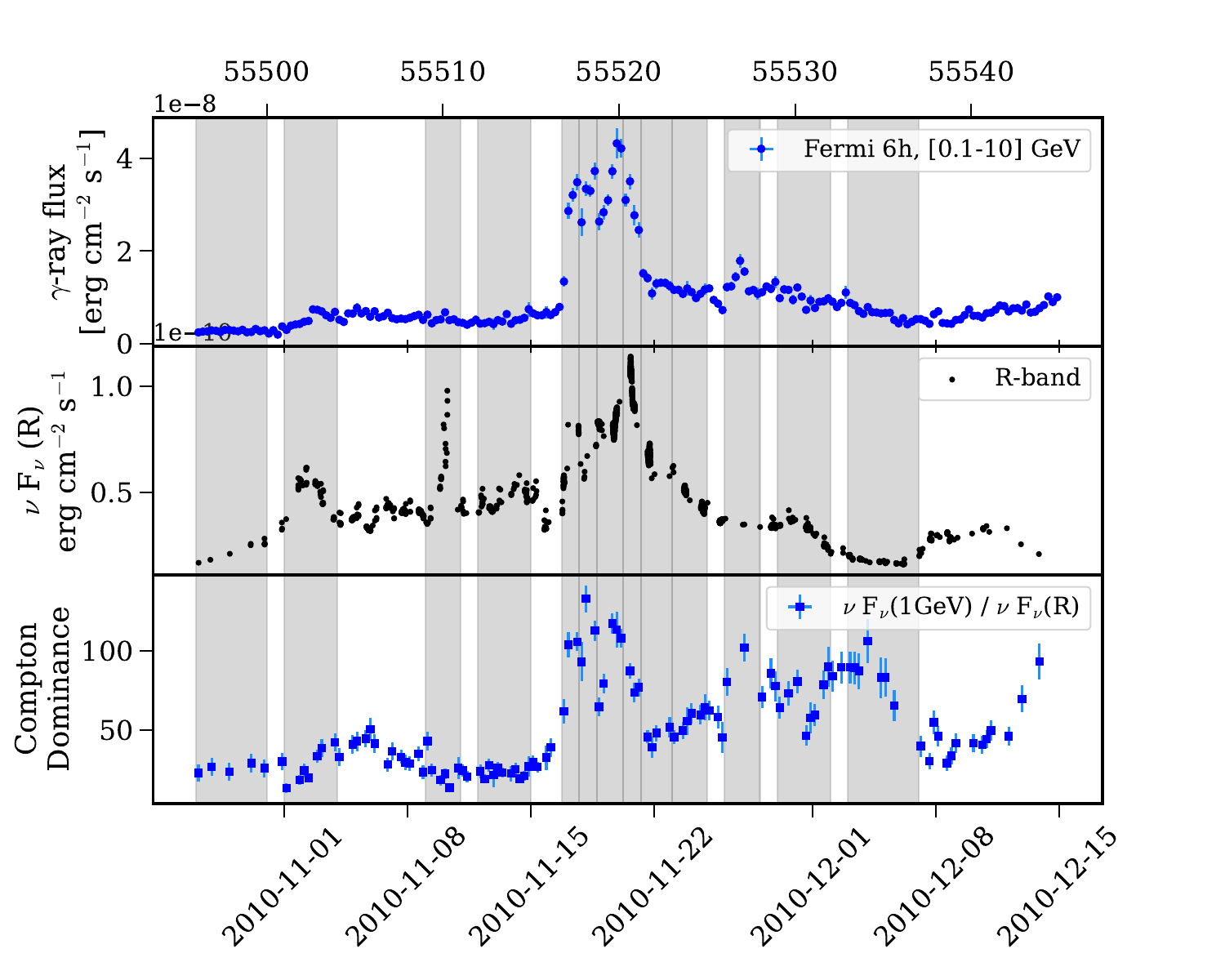}
        \caption{E10B, $6$ h}
        \label{fig:LightCurves_flaring_episodes:sub:10B}
    \end{subfigure}
    
    \begin{subfigure}{0.49\textwidth}
        \centering
        \includegraphics[width=\linewidth]{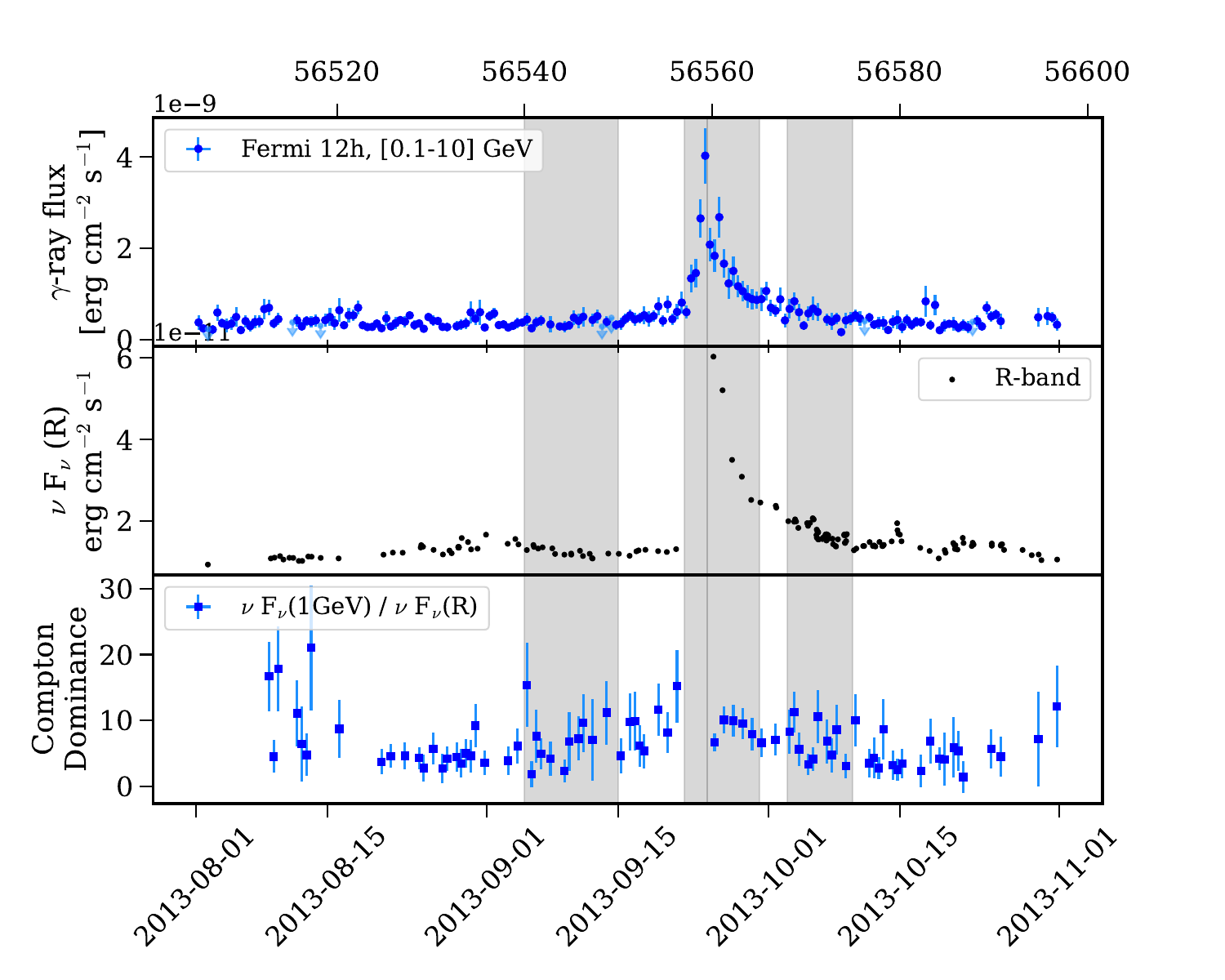}
        \caption{E13, 12 h}
        \label{fig:LightCurves_flaring_episodes:sub:13}
    \end{subfigure}
    \begin{subfigure}{0.49\textwidth}
        \centering
        \includegraphics[width=\linewidth]{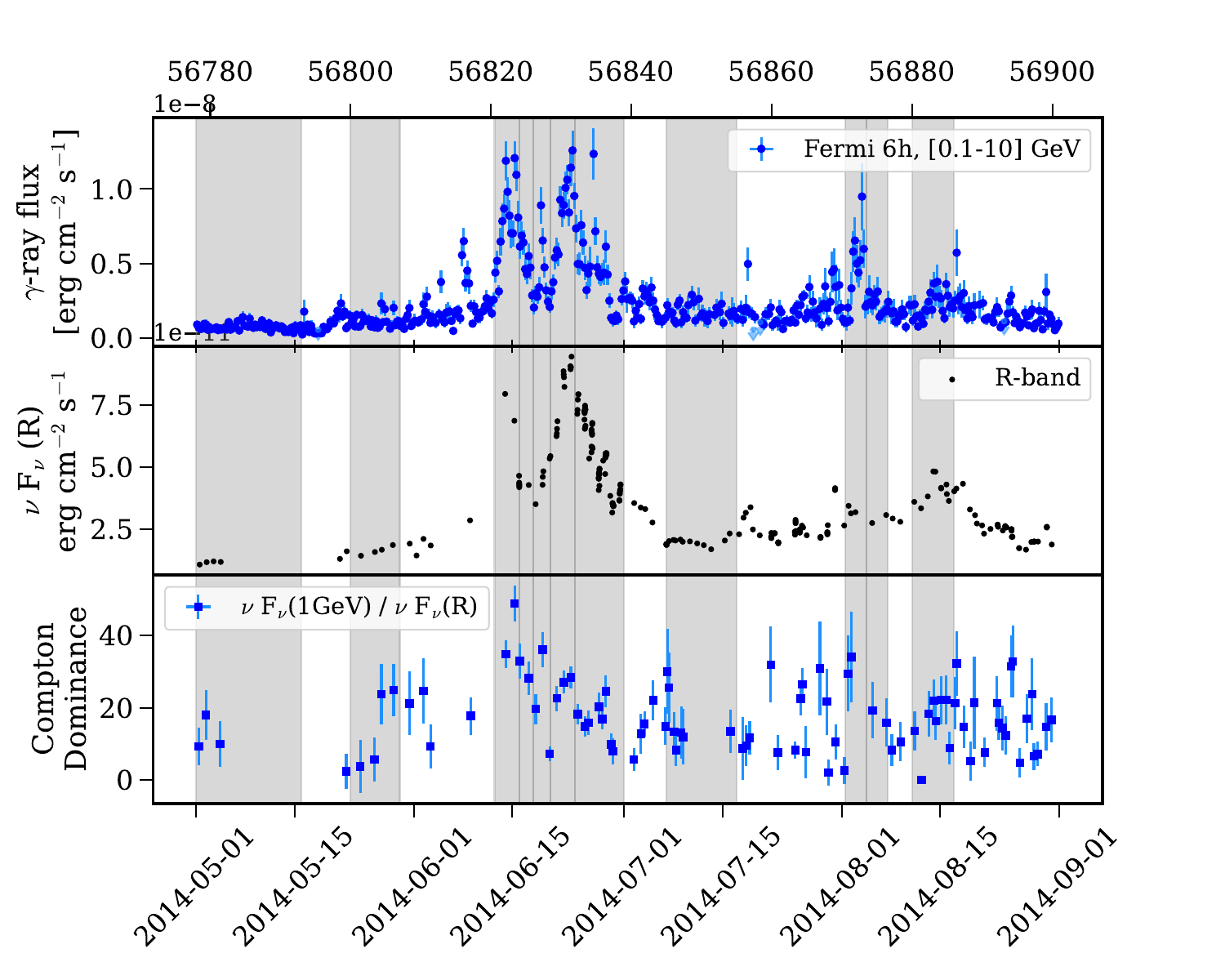}
        \caption{E14, $6$ h}
        \label{fig:LightCurves_flaring_episodes:sub:14}
    \end{subfigure}
    
    \caption{Continuation.}
    \label{fig:LightCurves_flaring_episodes_2}
\end{figure*}

\begin{figure*}
    \centering
    \ContinuedFloat

    \begin{subfigure}{0.49\textwidth}
        \centering
        \includegraphics[width=\linewidth]{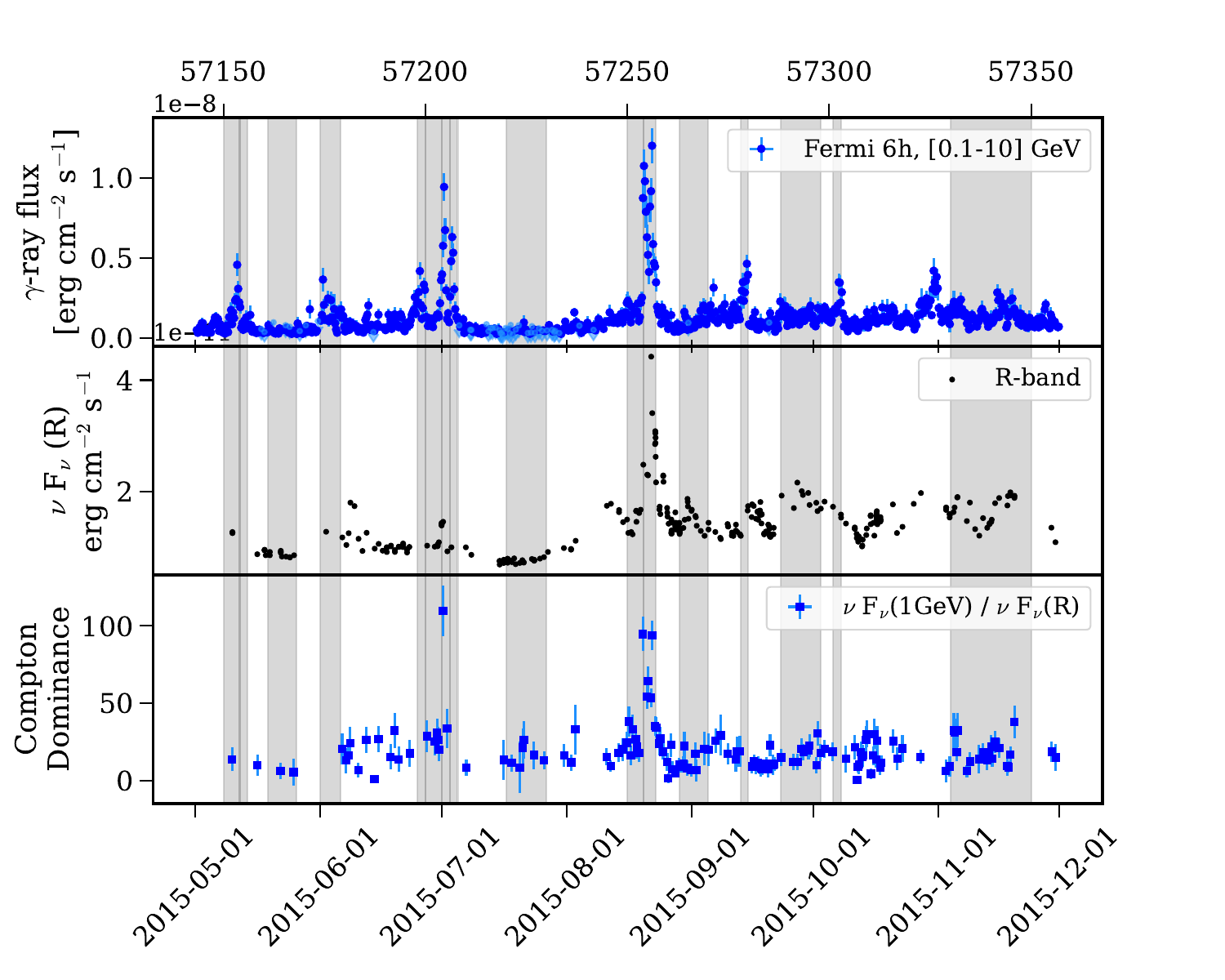}
        \caption{E15, $6$ h}
        \label{fig:LightCurves_flaring_episodes:sub:15}
    \end{subfigure}
    \begin{subfigure}{0.49\textwidth}
        \centering
        \includegraphics[width=\linewidth]{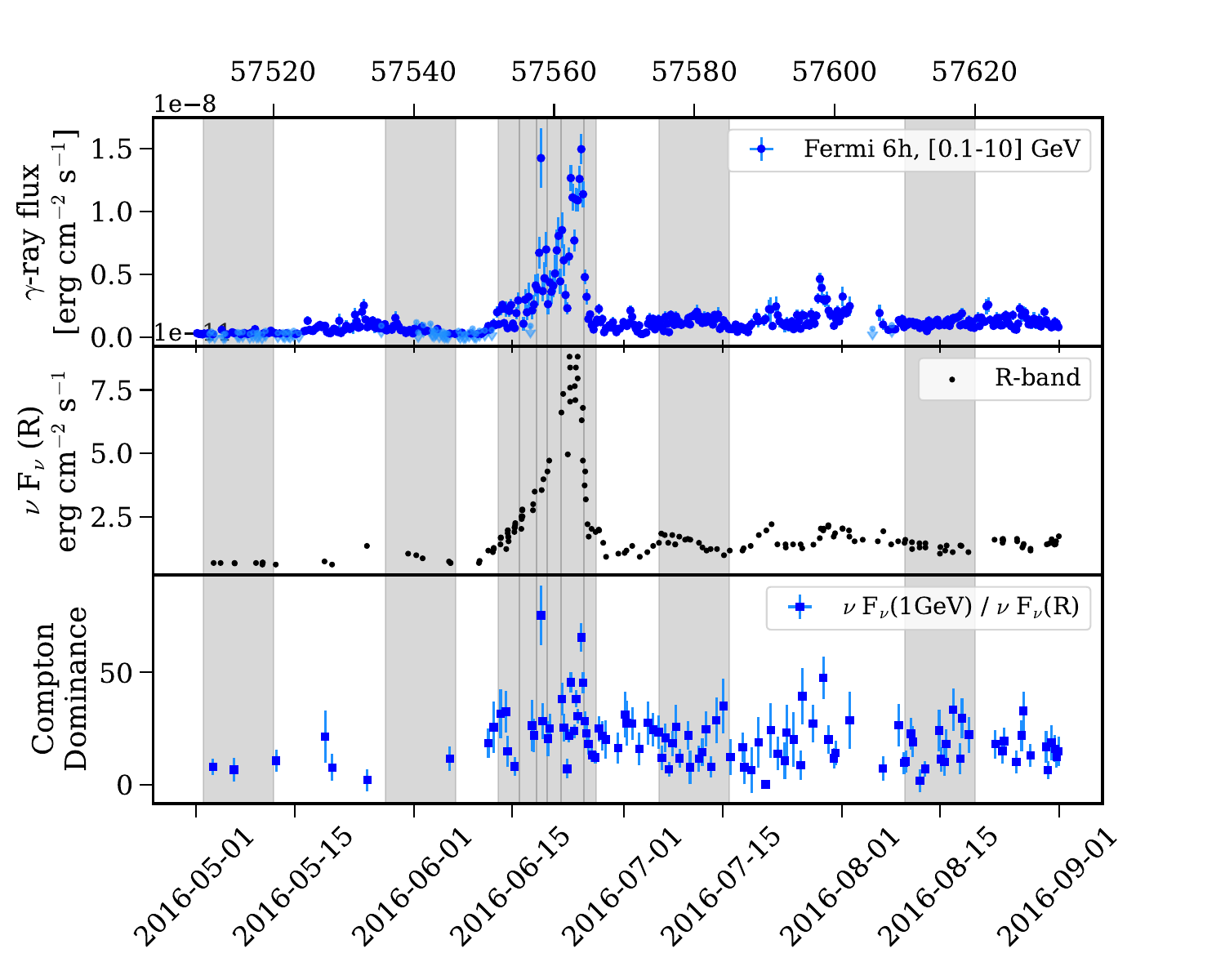}
        \caption{E16, $6$ h}
        \label{fig:LightCurves_flaring_episodes:sub:16}
    \end{subfigure}
    
    \begin{subfigure}{0.49\textwidth}
        \centering
        \includegraphics[width=\linewidth]{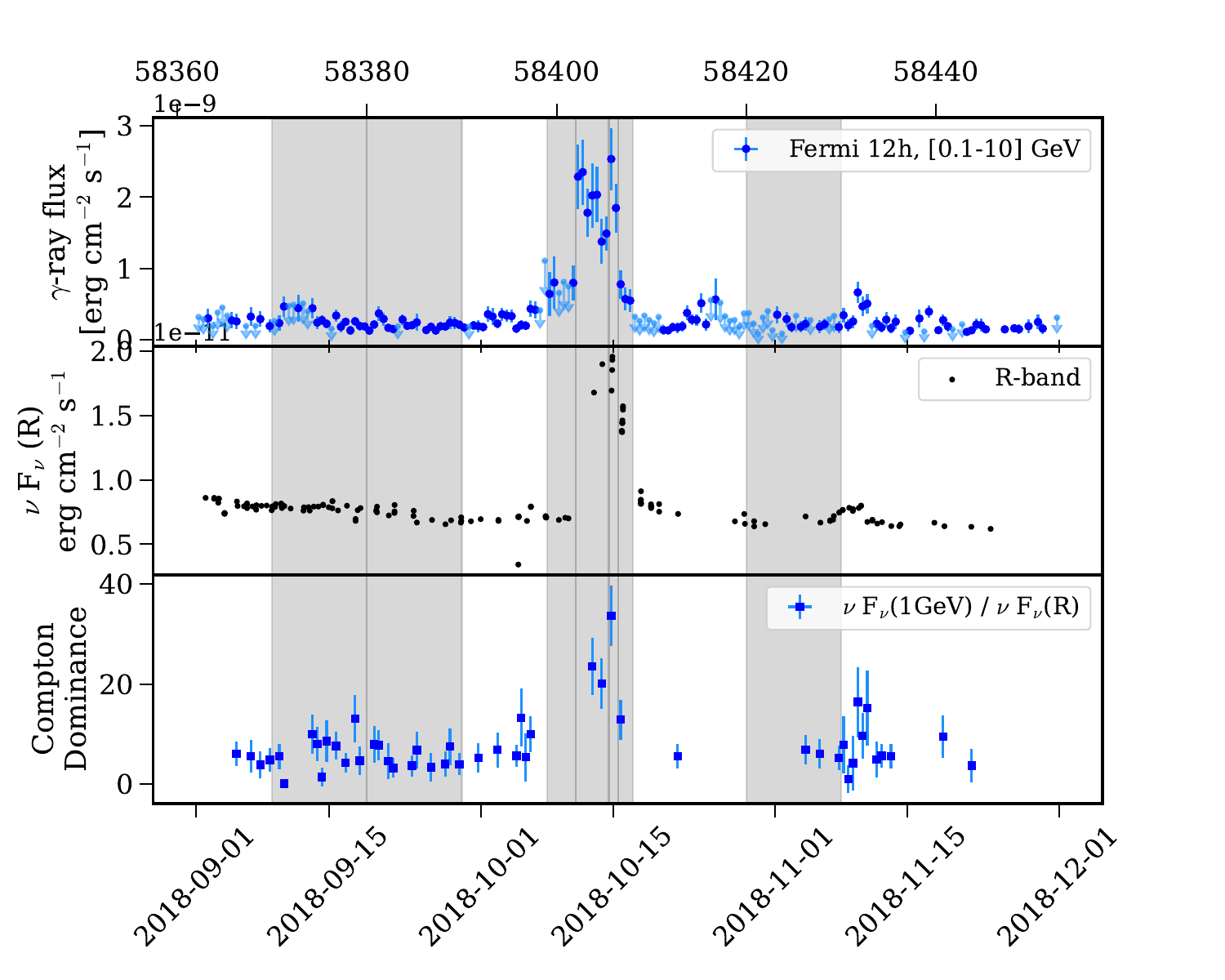}
        \caption{E18, $12$ h}
        \label{fig:LightCurves_flaring_episodes:sub:18}
    \end{subfigure}
    \begin{subfigure}{0.49\textwidth}
        \centering
        \includegraphics[width=\linewidth]{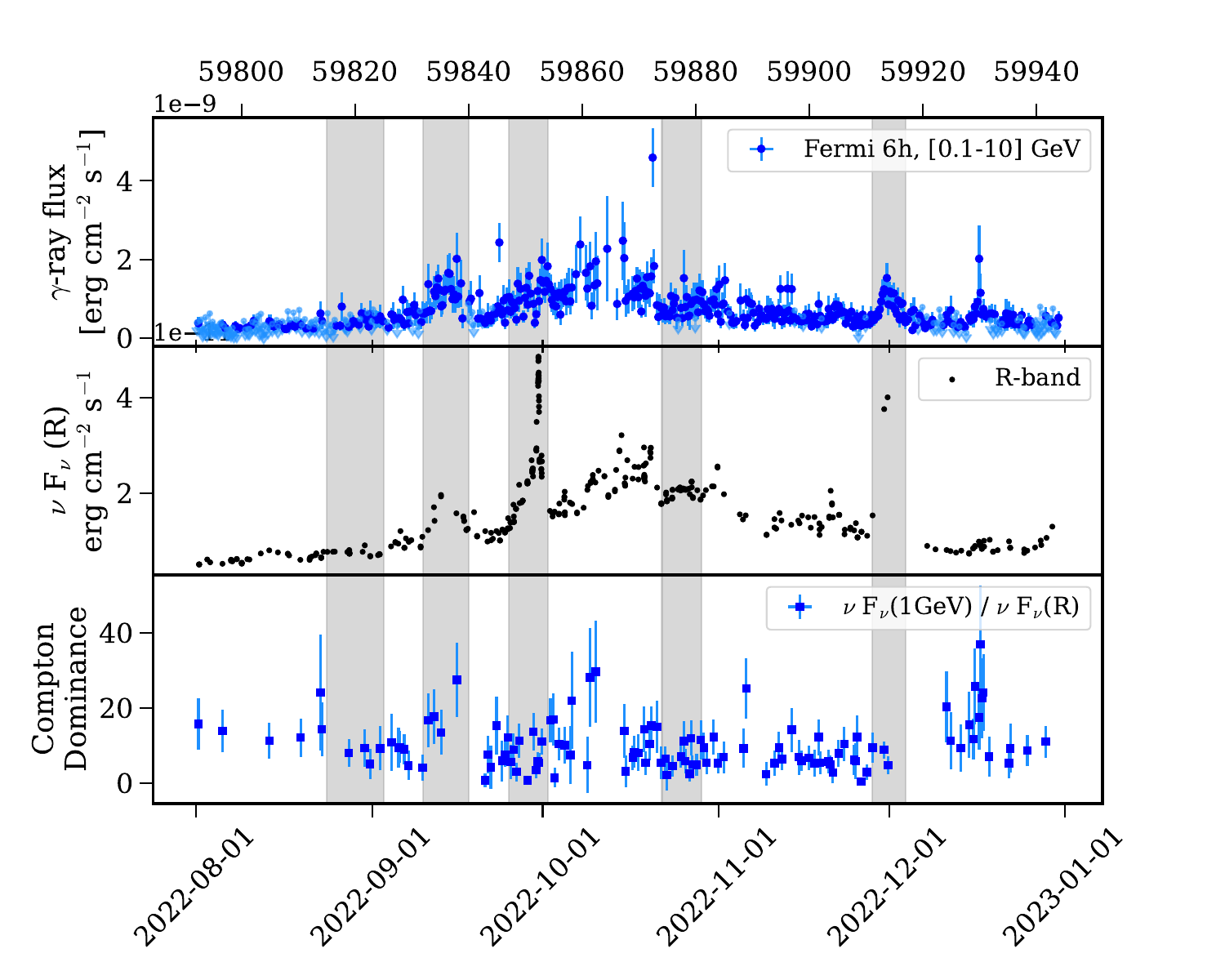}
        \caption{E22, $6$ h}
        \label{fig:LightCurves_flaring_episodes:sub:22}
    \end{subfigure}

    \caption{Continuation.}
    \label{fig:LightCurves_flaring_episodes_3}
    
\end{figure*}
%%%%%%%%%%%%%%%%%%%%%%%%%%%%%%%%%%%%%%%%%%%%%%%%%%%%%%%%%%%%%%%%%%%%%%%%%%%%%%%%%%%
%%%%%%%%%%%%%%%%%%%%%%%%%%%%%%%%%%%%%%%%%%%%%%%%%%%%%%%%%%%%%%%%%%%%%%%%%%%%%%%%%%%
%%%%%%%%%%%%%%%%%%%%%%%%%%%%%%%%%%%%%%%%%%%%%%%%%%%%%%%%%%%%%%%%%%%%%%%%%%%%%%%%%%%
%%%%%%%%%%%%%%%%%%%%%%%%%%%%%%%%%%%%%%%%%%%%%%%%%%%%%%%%%%%%%%%%%%%%%%%%%%%%%%%%%%%

%%%%%%%%%%%%%%%%%%%%%%%%%%%%%%%%%%%%%%%%%%
\begin{figure}
   \centering
   \resizebox{0.6\hsize}{!}{\includegraphics{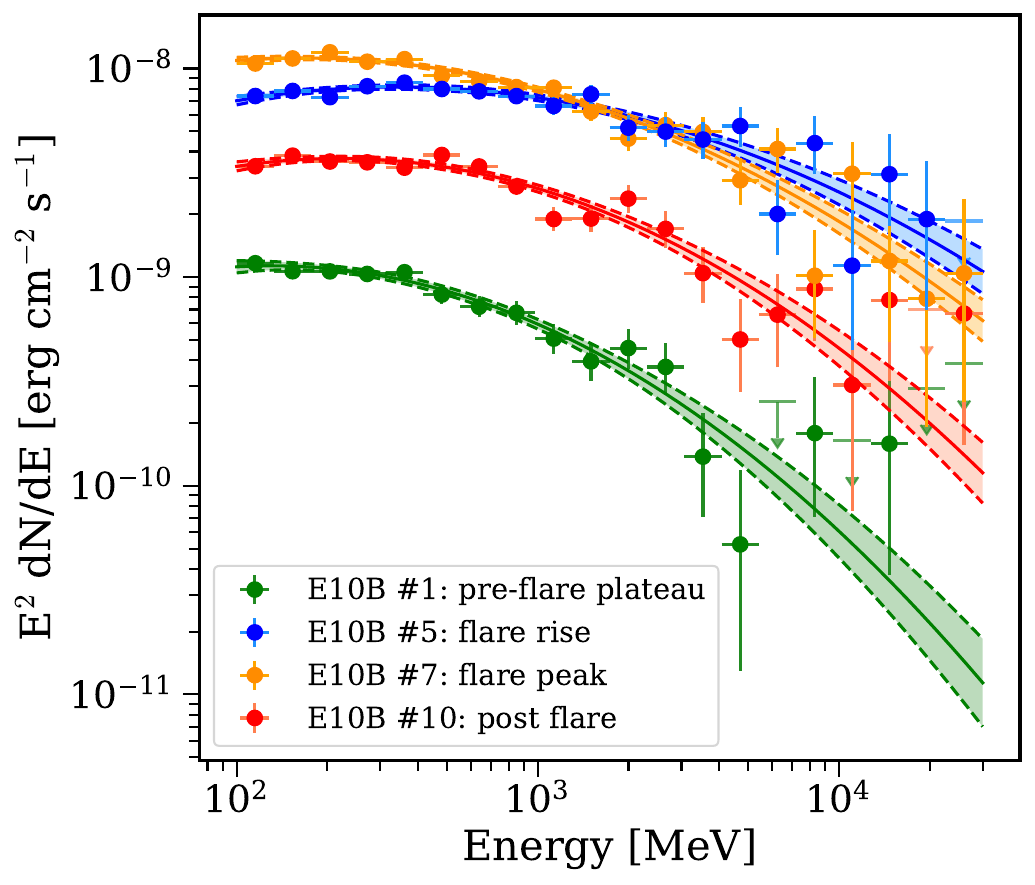}}
   \caption{Spectral evolution of epoch E10B, covering the brightest $\gamma$-ray flare of 3C~454.3. Best-fit parameters are listed in Table~\ref{table:spectra}.}
 \label{fig:spectra:2010B}
\end{figure}
%%%%%%%%%%%%%%%%%%%%%%%%%%%%%%%%%%%%%%%%%%

%%%%%%%%%%%%%%%%%%%%%%%%%%%%%%%%%%%%%%%%%%
\begin{figure}
   \centering
   \resizebox{0.6\hsize}{!}{\includegraphics{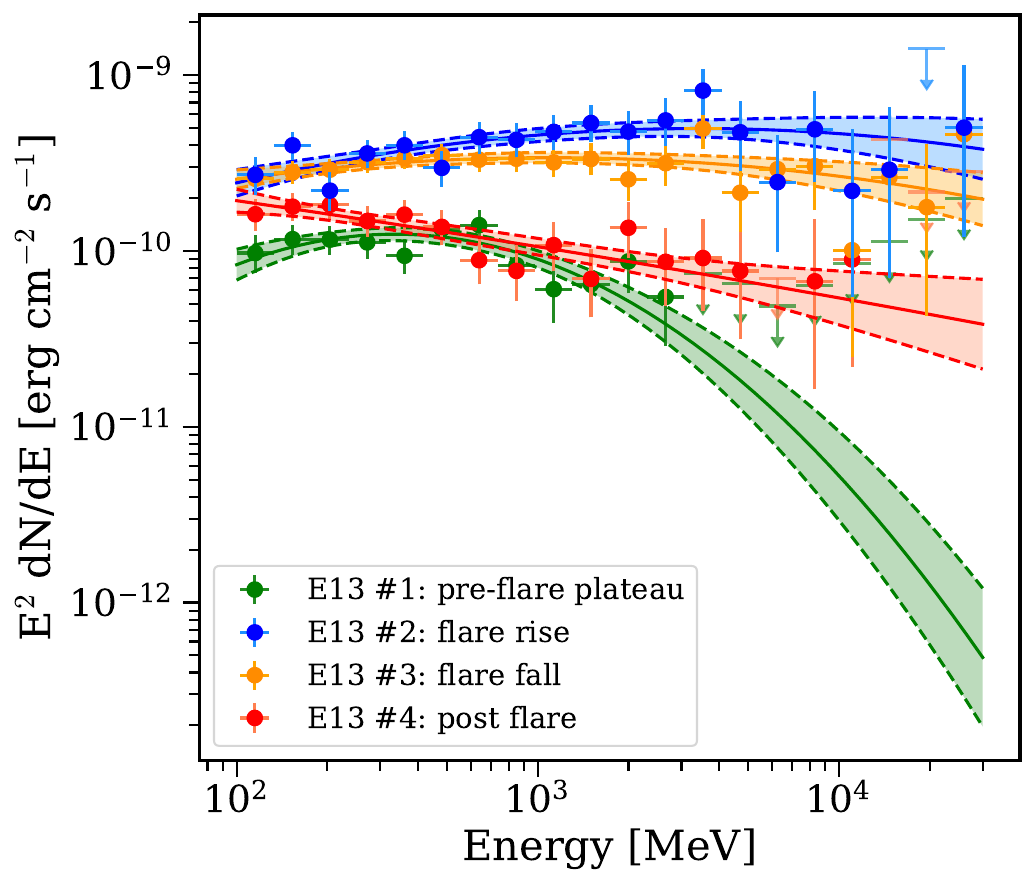}}
    \caption{Spectral evolution of epoch E13, showing significant hardening of the $\gamma$-ray spectrum during the flare. Best-fit parameters are listed in Table~\ref{table:spectra}.}
 \label{fig:spectra:2013}
\end{figure}
%%%%%%%%%%%%%%%%%%%%%%%%%%%%%%%%%%%%%%%%%%

\longtab[1]{
\begin{longtable}{rrr|rr|rrr}
\caption{
Best-fit spectral parameters of the $\gamma$-ray emission of 3C~454.3 during the selected temporal intervals. These integration periods correspond to the grey vertical bands shown in Fig.~\ref{fig:LightCurves_flaring_episodes}. Quiescent or low-activity states not listed in Table~\ref{table:Periods} are designated as `Q'. Columns: (1) start time in MJD; (2) stop time in MJD; (3) total integration duration $\Delta t$ in days; (4) parent flaring epoch from Table~\ref{table:Periods}; (5) specific identifier within the epoch; (6) log-parabola differential flux normalisation $N_0$; (7) spectral index $\alpha$; (8) curvature parameter $\beta$.
}
\label{table:spectra}
\\
\hline\hline
Start & Stop & $\Delta t$ & Epoch & ID & $N_0$                       & $\alpha$ & $\beta$ \\
MJD   & MJD  & d          &       &    & $cm^{-2}\,s^{-1}\,MeV^{-1}$ &          &         \\
\hline
\endfirsthead
\caption{continued.}\\
\hline\hline
Start & Stop & $\Delta t$ & Flare & ID & $N_0$                       & $\alpha$ & $\beta$ \\
MJD   & MJD  & d          &       &    & $cm^{-2}\,s^{-1}\,MeV^{-1}$ &          &         \\
\hline
\endhead
\hline
\endfoot
54694.0 & 54699.0 & 5.0 & E08B & 1 & $(2.5 \pm 0.1) \cdot 10^{-10}$ & $2.37\pm0.06$ & $0.10\pm0.03$ \\
54703.0 & 54704.5 & 1.5 & E08B & 2 & $(3.9 \pm 0.3) \cdot 10^{-10}$ & $2.44\pm0.09$ & $0.09\pm0.05$ \\
54706.0 & 54707.5 & 1.5 & E08B & 3 & $(4.5 \pm 0.4) \cdot 10^{-10}$ & $2.49\pm0.10$ & $0.20\pm0.06$ \\
54711.0 & 54715.0 & 4.0 & E08B & 4 & $(1.2 \pm 0.1) \cdot 10^{-10}$ & $2.46\pm0.10$ & $-0.02\pm0.05$ \\
54716.5 & 54719.5 & 3.0 & E08B & 5 & $(4.2 \pm 0.3) \cdot 10^{-10}$ & $2.41\pm0.08$ & $0.11\pm0.04$ \\
54721.0 & 54725.5 & 4.5 & E08B & 6 & $(3.7 \pm 0.2) \cdot 10^{-10}$ & $2.53\pm0.07$ & $0.14\pm0.04$ \\
54737.0 & 54740.0 & 3.0 & E08B & 7 & $(2.2 \pm 0.2) \cdot 10^{-10}$ & $2.51\pm0.09$ & $0.10\pm0.05$ \\
54740.0 & 54746.0 & 6.0 & E08B & 8 & $(2.3 \pm 0.1) \cdot 10^{-10}$ & $2.41\pm0.05$ & $0.11\pm0.03$ \\
54747.0 & 54750.0 & 3.0 & E08B & 9 & $(1.9 \pm 0.2) \cdot 10^{-10}$ & $2.34\pm0.08$ & $0.11\pm0.05$ \\
54771.0 & 54796.0 & 25.0 & E08B & 10 & $(1.07 \pm 0.05) \cdot 10^{-10}$ & $2.63\pm0.06$ & $0.16\pm0.03$ \\
54803.0 & 54818.0 & 15.0 & E08B & 11 & $(6.9 \pm 0.5) \cdot 10^{-11}$ & $2.64\pm0.08$ & $0.11\pm0.04$ \\
\hline
55058.0 & 55065.5 & 7.5 & E09A & 1 & $(9.9 \pm 0.8) \cdot 10^{-11}$ & $2.60\pm0.10$ & $0.16\pm0.06$ \\
55067.5 & 55071.0 & 3.5 & E09A & 2 & $(2.0 \pm 0.2) \cdot 10^{-10}$ & $2.49\pm0.08$ & $0.10\pm0.05$ \\
55071.5 & 55074.0 & 2.5 & E09A & 3 & $(2.7 \pm 0.2) \cdot 10^{-10}$ & $2.33\pm0.07$ & $0.03\pm0.04$ \\
55074.5 & 55077.5 & 3.0 & E09A & 4 & $(2.2 \pm 0.2) \cdot 10^{-10}$ & $2.42\pm0.09$ & $0.10\pm0.05$ \\
55079.0 & 55083.0 & 4.0 & E09A & 5 & $(1.6 \pm 0.2) \cdot 10^{-10}$ & $2.44\pm0.10$ & $0.11\pm0.07$ \\
55087.0 & 55091.5 & 4.5 & E09A & 6 & $(3.7 \pm 0.2) \cdot 10^{-10}$ & $2.45\pm0.06$ & $0.04\pm0.03$ \\
55091.5 & 55093.0 & 1.5 & E09A & 7 & $(3.5 \pm 0.4) \cdot 10^{-10}$ & $2.39\pm0.10$ & $0.06\pm0.07$ \\
55095.0 & 55098.0 & 3.0 & E09A & 8 & $(2.1 \pm 0.2) \cdot 10^{-10}$ & $2.48\pm0.10$ & $0.10\pm0.06$ \\
\hline
55140.0 & 55150.0 & 10.0 & E09B & 1 & $(1.7 \pm 0.1) \cdot 10^{-10}$ & $2.62\pm0.08$ & $0.17\pm0.04$ \\
55154.25 & 55155.75 & 1.5 & E09B & 2 & $(6.1 \pm 0.6) \cdot 10^{-10}$ & $2.33\pm0.09$ & $0.15\pm0.06$ \\
55160.5 & 55164.0 & 3.5 & E09B & 3 & $(7.8 \pm 0.4) \cdot 10^{-10}$ & $2.37\pm0.05$ & $0.14\pm0.03$ \\
55166.0 & 55168.0 & 2.0 & E09B & 4 & $(1.23 \pm 0.06) \cdot 10^{-9}$ & $2.43\pm0.05$ & $0.11\pm0.03$ \\
55168.0 & 55169.0 & 1.0 & E09B & 5 & $(1.39 \pm 0.08) \cdot 10^{-9}$ & $2.38\pm0.06$ & $0.09\pm0.03$ \\
55169.0 & 55170.5 & 1.5 & E09B & 6 & $(1.53 \pm 0.07) \cdot 10^{-9}$ & $2.34\pm0.04$ & $0.09\pm0.03$ \\
55171.0 & 55176.0 & 5.0 & E09B & 7 & $(1.11 \pm 0.03) \cdot 10^{-9}$ & $2.42\pm0.03$ & $0.10\pm0.02$ \\
55180.0 & 55189.0 & 9.0 & E09B & 8 & $(6.0 \pm 0.2) \cdot 10^{-10}$ & $2.37\pm0.03$ & $0.11\pm0.02$ \\
\hline
55250.0 & 55270.0 & 20.0 & E10A & 1 & $(2.04 \pm 0.08) \cdot 10^{-10}$ & $2.63\pm0.05$ & $0.13\pm0.03$ \\
55281.0 & 55287.0 & 6.0 & E10A & 2 & $(5.1 \pm 0.2) \cdot 10^{-10}$ & $2.47\pm0.05$ & $0.15\pm0.03$ \\
55287.0 & 55295.5 & 8.5 & E10A & 3 & $(9.9 \pm 0.2) \cdot 10^{-10}$ & $2.45\pm0.02$ & $0.10\pm0.01$ \\
55295.5 & 55298.0 & 2.5 & E10A & 4 & $(9.4 \pm 0.5) \cdot 10^{-10}$ & $2.45\pm0.06$ & $0.12\pm0.03$ \\
55299.0 & 55302.0 & 3.0 & E10A & 5 & $(1.01 \pm 0.05) \cdot 10^{-9}$ & $2.37\pm0.04$ & $0.05\pm0.02$ \\
55303.0 & 55310.0 & 7.0 & E10A & 6 & $(6.1 \pm 0.2) \cdot 10^{-10}$ & $2.54\pm0.05$ & $0.14\pm0.02$ \\
55315.0 & 55324.0 & 9.0 & E10A & 7 & $(4.4 \pm 0.2) \cdot 10^{-10}$ & $2.48\pm0.04$ & $0.10\pm0.02$ \\
55330.0 & 55340.0 & 10.0 & E10A & 8 & $(3.1 \pm 0.1) \cdot 10^{-10}$ & $2.42\pm0.04$ & $0.11\pm0.02$ \\
55372.0 & 55392.0 & 20.0 & E10A & 9 & $(1.19 \pm 0.07) \cdot 10^{-10}$ & $2.60\pm0.07$ & $0.17\pm0.04$ \\
\hline
55496.0 & 55500.0 & 4.0 & E10B & 1 & $(4.8 \pm 0.2) \cdot 10^{-10}$ & $2.60\pm0.06$ & $0.16\pm0.04$ \\
55501.0 & 55504.0 & 3.0 & E10B & 2 & $(1.01 \pm 0.04) \cdot 10^{-9}$ & $2.27\pm0.04$ & $0.10\pm0.02$ \\
55509.0 & 55511.0 & 2.0 & E10B & 3 & $(1.01 \pm 0.05) \cdot 10^{-9}$ & $2.43\pm0.06$ & $0.13\pm0.03$ \\
55512.0 & 55515.0 & 3.0 & E10B & 4 & $(9.2 \pm 0.4) \cdot 10^{-10}$ & $2.41\pm0.05$ & $0.11\pm0.03$ \\
55516.75 & 55517.75 & 1.0 & E10B & 5 & $(5.6 \pm 0.2) \cdot 10^{-9}$ & $2.20\pm0.03$ & $0.10\pm0.02$ \\
55517.75 & 55518.75 & 1.0 & E10B & 6 & $(6.1 \pm 0.2) \cdot 10^{-9}$ & $2.35\pm0.03$ & $0.10\pm0.02$ \\
55518.75 & 55520.25 & 1.5 & E10B & 7 & $(6.2 \pm 0.2) \cdot 10^{-9}$ & $2.37\pm0.03$ & $0.11\pm0.02$ \\
55520.25 & 55521.25 & 1.0 & E10B & 8 & $(5.8 \pm 0.2) \cdot 10^{-9}$ & $2.19\pm0.03$ & $0.12\pm0.02$ \\
55521.25 & 55523.0 & 1.75 & E10B & 9 & $(2.48 \pm 0.09) \cdot 10^{-9}$ & $2.38\pm0.04$ & $0.12\pm0.02$ \\
55523.0 & 55525.0 & 2.0 & E10B & 10 & $(2.10 \pm 0.08) \cdot 10^{-9}$ & $2.41\pm0.04$ & $0.14\pm0.02$ \\
55526.0 & 55528.0 & 2.0 & E10B & 11 & $(2.40 \pm 0.09) \cdot 10^{-9}$ & $2.44\pm0.04$ & $0.14\pm0.02$ \\
55529.0 & 55532.0 & 3.0 & E10B & 12 & $(1.78 \pm 0.06) \cdot 10^{-9}$ & $2.41\pm0.03$ & $0.11\pm0.02$ \\
55533.0 & 55537.0 & 4.0 & E10B & 13 & $(1.06 \pm 0.04) \cdot 10^{-9}$ & $2.49\pm0.04$ & $0.10\pm0.02$ \\
\hline
55700.0 & 55900.0 & 200.0 & Q & 1 & $(1.26 \pm 0.08) \cdot 10^{-11}$ & $2.65\pm0.09$ & $0.19\pm0.05$ \\
\hline
56540.0 & 56550.0 & 10.0 & E13 & 1 & $(7.2 \pm 0.8) \cdot 10^{-11}$ & $2.55\pm0.10$ & $0.27\pm0.09$ \\
56557.0 & 56559.5 & 2.5 & E13 & 2 & $(3.4 \pm 0.3) \cdot 10^{-10}$ & $1.85\pm0.06$ & $0.06\pm0.04$ \\
56559.5 & 56565.0 & 5.5 & E13 & 3 & $(2.6 \pm 0.2) \cdot 10^{-10}$ & $1.98\pm0.04$ & $0.05\pm0.03$ \\
56568.0 & 56575.0 & 7.0 & E13 & 4 & $(8.2 \pm 0.9) \cdot 10^{-11}$ & $2.28\pm0.08$ & $0.01\pm0.05$ \\
\hline
56778.0 & 56793.0 & 15.0 & E14 & 1 & $(1.27 \pm 0.07) \cdot 10^{-10}$ & $2.58\pm0.07$ & $0.17\pm0.04$ \\
56800.0 & 56807.0 & 7.0 & E14 & 2 & $(1.9 \pm 0.1) \cdot 10^{-10}$ & $2.36\pm0.06$ & $0.10\pm0.04$ \\
56820.5 & 56824.0 & 3.5 & E14 & 3 & $(1.53 \pm 0.05) \cdot 10^{-9}$ & $2.05\pm0.03$ & $0.10\pm0.02$ \\
56824.0 & 56826.0 & 2.0 & E14 & 4 & $(9.9 \pm 0.6) \cdot 10^{-10}$ & $2.04\pm0.04$ & $0.10\pm0.03$ \\
56826.0 & 56828.5 & 2.5 & E14 & 5 & $(7.5 \pm 0.4) \cdot 10^{-10}$ & $2.04\pm0.04$ & $0.09\pm0.03$ \\
56828.5 & 56832.0 & 3.5 & E14 & 6 & $(1.53 \pm 0.05) \cdot 10^{-9}$ & $1.98\pm0.02$ & $0.09\pm0.02$ \\
56832.0 & 56839.0 & 7.0 & E14 & 7 & $(7.7 \pm 0.3) \cdot 10^{-10}$ & $1.99\pm0.02$ & $0.07\pm0.02$ \\
56845.0 & 56855.0 & 10.0 & E14 & 8 & $(3.4 \pm 0.2) \cdot 10^{-10}$ & $2.35\pm0.06$ & $0.13\pm0.03$ \\
56870.5 & 56873.5 & 3.0 & E14 & 9 & $(7.8 \pm 0.7) \cdot 10^{-10}$ & $2.01\pm0.06$ & $0.05\pm0.04$ \\
56873.5 & 56876.5 & 3.0 & E14 & 10 & $(3.9 \pm 0.5) \cdot 10^{-10}$ & $2.29\pm0.10$ & $0.10\pm0.07$ \\
56880.0 & 56886.0 & 6.0 & E14 & 11 & $(3.6 \pm 0.3) \cdot 10^{-10}$ & $2.48\pm0.08$ & $0.11\pm0.05$ \\
\hline
57150.0 & 57154.0 & 4.0 & E15 & 1 & $(2.8 \pm 0.2) \cdot 10^{-10}$ & $2.14\pm0.06$ & $0.10\pm0.04$ \\
57154.0 & 57156.0 & 2.0 & E15 & 2 & $(1.5 \pm 0.2) \cdot 10^{-10}$ & $2.20\pm0.10$ & $0.02\pm0.07$ \\
57161.0 & 57168.0 & 7.0 & E15 & 3 & $(7.4 \pm 0.9) \cdot 10^{-11}$ & $2.66\pm0.20$ & $0.21\pm0.09$ \\
57174.0 & 57179.0 & 5.0 & E15 & 4 & $(3.0 \pm 0.2) \cdot 10^{-10}$ & $2.13\pm0.06$ & $0.14\pm0.04$ \\
57198.0 & 57200.0 & 2.0 & E15 & 5 & $(4.9 \pm 0.4) \cdot 10^{-10}$ & $2.28\pm0.07$ & $0.17\pm0.04$ \\
57200.0 & 57204.0 & 4.0 & E15 & 6 & $(2.2 \pm 0.2) \cdot 10^{-10}$ & $2.47\pm0.08$ & $0.09\pm0.05$ \\
57204.0 & 57206.0 & 2.0 & E15 & 7 & $(7.7 \pm 0.5) \cdot 10^{-10}$ & $2.41\pm0.07$ & $0.16\pm0.04$ \\
57206.0 & 57208.0 & 2.0 & E15 & 8 & $(6.4 \pm 0.5) \cdot 10^{-10}$ & $2.46\pm0.08$ & $0.18\pm0.05$ \\
57220.0 & 57230.0 & 10.0 & E15 & 9 & $(6.3 \pm 0.7) \cdot 10^{-11}$ & $2.63\pm0.10$ & $0.21\pm0.08$ \\
57250.0 & 57254.0 & 4.0 & E15 & 10 & $(4.3 \pm 0.3) \cdot 10^{-10}$ & $2.26\pm0.06$ & $0.15\pm0.04$ \\
57254.0 & 57257.0 & 3.0 & E15 & 11 & $(1.45 \pm 0.06) \cdot 10^{-9}$ & $2.19\pm0.03$ & $0.11\pm0.02$ \\
57263.0 & 57270.0 & 7.0 & E15 & 12 & $(1.7 \pm 0.1) \cdot 10^{-10}$ & $2.55\pm0.09$ & $0.18\pm0.05$ \\
57278.0 & 57280.0 & 2.0 & E15 & 13 & $(6.1 \pm 0.5) \cdot 10^{-10}$ & $2.17\pm0.06$ & $0.08\pm0.04$ \\
57288.0 & 57298.0 & 10.0 & E15 & 14 & $(2.4 \pm 0.1) \cdot 10^{-10}$ & $2.53\pm0.06$ & $0.15\pm0.03$ \\
57301.0 & 57303.0 & 2.0 & E15 & 15 & $(4.4 \pm 0.4) \cdot 10^{-10}$ & $2.44\pm0.09$ & $0.18\pm0.06$ \\
57330.0 & 57350.0 & 20.0 & E15 & 16 & $(2.53 \pm 0.09) \cdot 10^{-10}$ & $2.40\pm0.04$ & $0.13\pm0.02$ \\
\hline
57510.0 & 57520.0 & 10.0 & E16 & 1 & $(3.7 \pm 0.5) \cdot 10^{-11}$ & $2.71\pm0.10$ & $0.09\pm0.08$ \\
57536.0 & 57546.0 & 10.0 & E16 & 2 & $(7.3 \pm 0.7) \cdot 10^{-11}$ & $2.52\pm0.10$ & $0.10\pm0.06$ \\
57552.0 & 57555.0 & 3.0 & E16 & 3 & $(3.6 \pm 0.3) \cdot 10^{-10}$ & $2.23\pm0.07$ & $0.12\pm0.04$ \\
57555.0 & 57557.5 & 2.5 & E16 & 4 & $(4.8 \pm 0.6) \cdot 10^{-10}$ & $2.13\pm0.10$ & $0.16\pm0.07$ \\
57557.5 & 57559.0 & 1.5 & E16 & 5 & $(1.2 \pm 0.1) \cdot 10^{-9}$ & $1.99\pm0.07$ & $0.08\pm0.05$ \\
57559.0 & 57561.0 & 2.0 & E16 & 6 & $(1.02 \pm 0.09) \cdot 10^{-9}$ & $2.13\pm0.08$ & $0.16\pm0.06$ \\
57561.0 & 57564.25 & 3.25 & E16 & 7 & $(2.07 \pm 0.06) \cdot 10^{-9}$ & $2.15\pm0.03$ & $0.14\pm0.02$ \\
57564.25 & 57566.0 & 1.75 & E16 & 8 & $(2.9 \pm 0.2) \cdot 10^{-10}$ & $2.30\pm0.07$ & $0.09\pm0.04$ \\
57575.0 & 57585.0 & 10.0 & E16 & 9 & $(2.2 \pm 0.1) \cdot 10^{-10}$ & $2.41\pm0.05$ & $0.10\pm0.03$ \\
57610.0 & 57620.0 & 10.0 & E16 & 10 & $(2.0 \pm 0.1) \cdot 10^{-10}$ & $2.49\pm0.06$ & $0.14\pm0.04$ \\
\hline
57900.0 & 58000.0 & 100.0 & Q & 2 & $(2.3 \pm 0.1) \cdot 10^{-11}$ & $2.68\pm0.08$ & $0.16\pm0.05$ \\
\hline
58370.0 & 58380.0 & 10.0 & E18 & 1 & $(3.9 \pm 0.5) \cdot 10^{-11}$ & $2.79\pm0.20$ & $0.21\pm0.09$ \\
58380.0 & 58390.0 & 10.0 & E18 & 2 & $(3.0 \pm 0.3) \cdot 10^{-11}$ & $2.66\pm0.10$ & $0.10\pm0.08$ \\
58399.0 & 58402.0 & 3.0 & E18 & 3 & $(1.1 \pm 0.3) \cdot 10^{-10}$ & $2.74\pm0.40$ & $0.33\pm0.20$ \\
58402.0 & 58405.5 & 3.5 & E18 & 4 & $(3.9 \pm 0.3) \cdot 10^{-10}$ & $2.00\pm0.06$ & $0.16\pm0.05$ \\
58405.5 & 58406.5 & 1.0 & E18 & 5 & $(4.1 \pm 0.5) \cdot 10^{-10}$ & $1.81\pm0.10$ & $0.20\pm0.08$ \\
58406.5 & 58408.0 & 1.5 & E18 & 6 & $(1.0 \pm 0.2) \cdot 10^{-10}$ & $2.15\pm0.20$ & $0.09\pm0.10$ \\
58420.0 & 58430.0 & 10.0 & E18 & 7 & $(2.4 \pm 0.5) \cdot 10^{-11}$ & $2.69\pm0.20$ & $0.15\pm0.10$ \\
\hline
58800.0 & 59200.0 & 400.0 & Q & 3 & $(1.27 \pm 0.06) \cdot 10^{-11}$ & $2.60\pm0.05$ & $0.16\pm0.03$ \\
\hline
59815.0 & 59825.0 & 10.0 & E22 & 1 & $(5.6 \pm 0.7) \cdot 10^{-11}$ & $2.49\pm0.10$ & $0.07\pm0.06$ \\
59832.0 & 59840.0 & 8.0 & E22 & 2 & $(1.9 \pm 0.1) \cdot 10^{-10}$ & $2.35\pm0.07$ & $0.08\pm0.04$ \\
59847.0 & 59854.0 & 7.0 & E22 & 3 & $(1.8 \pm 0.1) \cdot 10^{-10}$ & $2.33\pm0.06$ & $0.08\pm0.04$ \\
59874.0 & 59881.0 & 7.0 & E22 & 4 & $(1.2 \pm 0.1) \cdot 10^{-10}$ & $2.27\pm0.07$ & $0.01\pm0.04$ \\
59911.0 & 59917.0 & 6.0 & E22 & 5 & $(1.5 \pm 0.1) \cdot 10^{-10}$ & $2.52\pm0.09$ & $0.14\pm0.05$ \\
\end{longtable}
}

\end{appendix}

\end{document}